# Estimating the proportion of modern contraceptives supplied by the public and private sectors using a Bayesian hierarchical penalized spline model

Hannah Comiskey [a]*, Prof. Leontine Alkema [b], Dr. Niamh Cahill [a]

* Email: hannah.comiskey.2015@mumail.ie

[a]Department of Mathematics & Statistics, Maynooth University, Maynooth, Ireland; [b]School of Public Health & Health Sciences, University of Massachusetts Amherst, Massachusetts, USA

Quantifying the public/private sector supply of contraceptive methods within countries is vital for effective and sustainable family planning (FP) delivery. In many low and middle-income countries (LMIC), measuring the contraceptive supply source often relies on Demographic Health Surveys (DHS). However, many of these countries carry out the DHS approximately every 3-5 years and do not have recent data beyond 2015/16. Our objective in estimating the set of related contraceptive supply-share outcomes (proportion of modern contraceptive methods supplied by the public/private sectors) is to take advantage of latent attributes present in dataset to produce annual, country-specific estimates and projections with uncertainty. We propose a Bayesian, hierarchical, penalized-spline model with multivariate-normal spline coefficients to capture cross-method correlations. Our approach offers an intuitive way to share information across countries and sub-continents, model the changes in the contraceptive supply share over time, account for survey observational errors and produce probabilistic estimates and projections that are informed by past changes in the contraceptive supply share as well as correlations between rates of change across different methods. These results will provide valuable information for evaluating FP program effectiveness. To the best of our knowledge, it is the first model of its kind to estimate these quantities.

**Introduction**

The Family Planning 2020 Initiative (FP2020) set the target of engaging 120 million new users of modern contraception in sixty-nine of the world's poorest countries by the year 2020. This goal has now been extended to 2030 in line with the UN sustainable development goals[1,2]. Consequently, there is a heightened need for annual, reliable estimates of family planning (FP) indicators for each country involved in the initiative. One such indicator is the most recent source of contraceptive commodities accessed by FP users. This indicator can be used to highlight where FP users obtain their contraceptives from, as well as to evaluate FP program effectiveness and forecast future commodity procurement[3]. The family planning market is most successful when clients have a variety of methods and sources to choose from[4]. Therefore, annual estimates and projections of the contraceptive method supply share are of particular use to countries who are trying to engage the private sector to rebalance the roles of the public and private sectors[3], meet the needs of all women within a country and enhance efforts to meet family planning goals[5].

In addition to their role as a stand-alone FP indicator, estimates of the contraceptive method supply share can be utilised with FP service statistics in a corrective step to reduce the bias of estimated modern contraceptive use (EMUs) rates[6]. EMUs represent the proportion of women in a particular country using modern contraception based on FP service statistic data[7]. Service statistics data are routinely collected in connection with family planning service delivery. These data provide high geographic detail, are relatively inexpensive to collect but often do not include contributions from the private sector[8]. The private sector is used by women across all socio-economic groups to access their family planning methods[9], with 37% to 39% of all family planning users obtaining modern contraceptives through the private sector[10]. Therefore, the private sector makes up a significant portion of the family planning market. Service statistics data that are missing the private sector's contribution must be scaled-up to reflect the complete contraceptive market[11]. This adjustment to the service statistics data has spurred the need for annual estimates of the proportion of modern contraceptives supplied by the public, private commercial medical and private other sectors. Presently, annual estimates and projections with uncertainty for these proportions are not available.

DHS surveys provide information on the breakdown of public and private sector contributions to the contraceptive supply chain in the form of the percentage of contraceptive methods that are sourced from each sector. In this study, we consider 30 countries involved in the FP2020 initiative, that have DHS data available after 2012. Notably, 54% of the countries considered do not have survey observations of the supply share indicator beyond 2015. In the absence of recent survey observations we can rely on statistical model-based estimates and projections to bridge the knowledge gap and provide the recent private sector contributions to family planning service delivery with uncertainty.

We describe a Bayesian hierarchical penalised spline model that produces annual, country and method-specific estimates of the proportion of modern contraceptives coming from the public and private sectors. The model accounts for across method correlations within public and private sector contributions to family planning supply and relies on information sharing across methods and countries via a hierarchical modelling structure.

**Data**

*Definitions and Data Sources*

According to Hubacher and Trussell, a modern contraceptive method is defined as "a product or medical procedure that interferes with reproduction from acts of sexual intercourse"[12]. Modern methods of contraception considered in this study are female sterilisation, oral contraceptive pills (OC pills), implants (including Implanon, Jadelle and Sino-implant), intra-uterine devices (IUD, including Copper- T 380-A IUD and LNG-IUS), and injectables (including Depo Provera (DMPA), Noristerat (NET-En), Lunelle, Sayana Press and other injectables).

The proportion of a modern contraceptive method provided by the public sector is defined as the percent of a modern contraceptive method that is supplied by any public sector outlet relative to all modern methods supplied in each country at a given time. Conversely, the proportion of modern contraceptives provided by the private sector come from the private sector outlets[13]. The public sectors include contraceptives supplied by government health facilities and home/community deliveries. Any supplies that come from sources outside the public sector can be defined as coming from the private sector. These include commercial, for-profit, and non-profit organizations[14]. We consider three sector categories: public, private commercial medical and other private, where the private commercial medical and other private make up the total private sector. In this study, the outcome of interest for a given contraceptive method is the market share breakdown using the three sector categories.

A database of the public and private sector breakdown of modern contraceptive supply with their associated standard errors was created using data from the DHS[15]. The DHS use a two-stage sampling design, using census information as the sampling frame. First, the country of interest is stratified to have strata as homogenous as possible. This homogeneity minimizes the resulting sampling errors of the survey. Within each stratum, the census enumeration areas (EAs) form clusters. The households of each cluster are listed, and a fixed number of households within the selected cluster are chosen by systematic sampling. The sample selection process uses weights to address the any differences in probability of selection. This corrects any over- or under-sampling of different clusters during sample selection. The DHS standard model questionnaires are then utilised to collect data from the selected households[16]. The cohort of interest for this study was women aged 15-49 years old who are taking a modern contraceptive method. The variable of interest was the current source of their modern contraceptive methods.

In this study we consider countries involved in the FP2030 initiative. Table 1 lists the thirty countries used in this study. The total number of surveys carried out and the year of the most recent survey is listed for each country. Just under half of the countries included have survey data available after 2015, highlighting the need for annual up-to-date estimates of the contraceptive supply shares.

| Country | Total Number of Surveys | Recent Survey Year |
|:---:|:---:|:---:|
| Afghanistan | 1 | 2015 |
| Benin | 5 | 2017 |
| Burkina Faso | 4 | 2010 |
| Cameroon | 5 | 2018 |
| Congo | 1 | 2005 |
| Congo Democratic Republic | 2 | 2013 |
| Cote d'Ivoire | 3 | 2011 |
| Ethiopia | 5 | 2019 |
| Ghana | 5 | 2014 |
| Guinea | 4 | 2018 |
| India | 4 | 2005 |
| Kenya | 5 | 2014 |
| Liberia | 4 | 2019 |
| Madagascar | 4 | 2008 |
| Malawi | 5 | 2015 |
| Mali | 5 | 2018 |
| Mozambique | 3 | 2011 |
| Myanmar | 1 | 2015 |
| Nepal | 5 | 2016 |
| Niger | 4 | 2012 |
| Nigeria | 5 | 2018 |
| Pakistan | 4 | 2017 |
| Philippines | 6 | 2017 |
| Rwanda | 6 | 2019 |
| Senegal | 10 | 2019 |
| Sierra Leone | 3 | 2019 |
| Tanzania | 6 | 2015 |
| Togo | 2 | 2013 |
| Uganda | 5 | 2016 |
| Zimbabwe | 5 | 2015 |

**Table 1.** Summary of DHS microdata used during the study including the country names, the number of DHS surveys per country available and the year of the most recent DHS survey available. Just over 46% of countries have data available after 2015.

We calculated proportions and associated standard errors from DHS micro-level data. In line with STATcompiler[17] convention, we filtered the DHS survey microdata to only include observations for a given method $m$, at time $t$ in country $c$ where at least one sector (public, private commercial or private other) has a sample size of at least twenty women. This removes sets of observations with large uncertainty due to small sample sizes. Sampling errors were calculated while accounting for the sampling design using a Taylor series linearisation method to approximate the standard error of the calculated proportions[18,19].

   The final database included 1308 observations for 15 method-sector combinations. An example of the data is shown in Figure 1 where the proportion of modern contraceptives coming from each sector are plotted over time for Zimbabwe. Vertical bars indicate standard errors associated with each observation. Zimbabwe is considered a data-rich country as there are between 3 and 5 surveys available for each method. For all five methods, the public sector has the largest market share. However, in recent years, there is an increase in the proportion of OC pills and female sterilization supplied by the private sectors. In the database, standard errors range from 0.015 to 22.6 percentage points. For further details on how the standard errors were calculated and a summary of these errors, please see the Appendix, section 1.

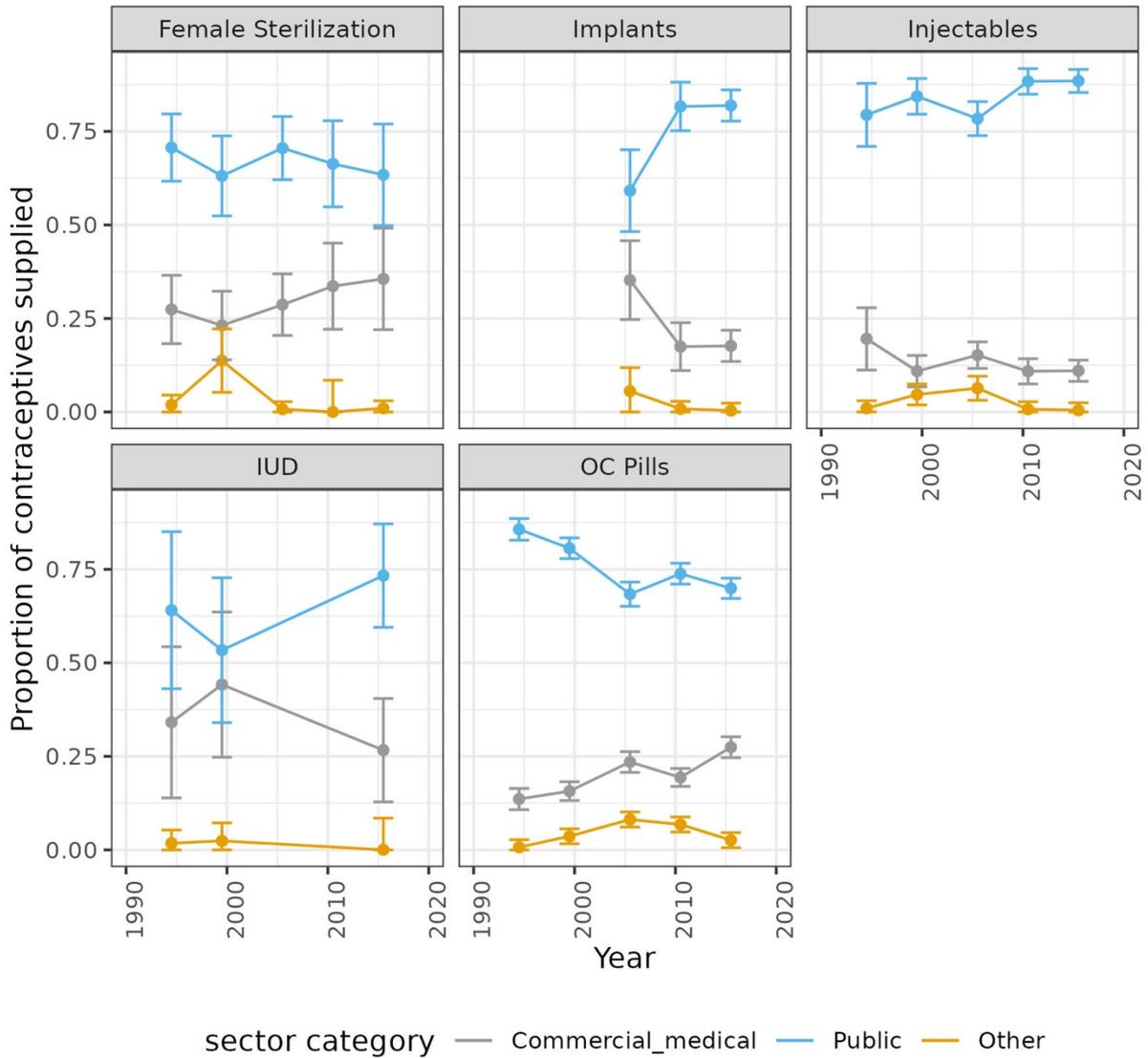

**Figure 1.** Observed proportions of the modern contraceptives supplied by each sector over time for Zimbabwe according to the DHS microdata. The sectors are coloured blue for public, grey for commercial medical and gold for other private. The standard errors associated with each observation are used to create the vertical error bars. Zimbabwe is one of the thirty countries included in this study.

**Methods**

The outcome of interest is the components of a compositional vector $\boldsymbol{\phi}_{c,t,m} = (\phi_{c,t,m,1}, \phi_{c,t,m,2}, \phi_{c,t,m,3})$ denoting $\phi_{c,t,m,s}$ the proportion supplied by the public sector ($s=1$), the private commercial medical sector ($s=2$) and the other private sector ($s=3$) of modern contraceptive method m, at time t, in country c.

We break the model specification into two parts, the process model that captures the underlying dynamics of the outcome of interest and the data model which links the observed data to the process model. In the process model, we model the logit-transformed proportion of the public-sector supply share and the ratio of private commercial medical to total private sector supply share using a Bayesian hierarchical penalised spline model. The data is linked to the process in the data model via a truncated normal distribution.

***The process model***

We begin by defining a regression model for $\phi_{c,t,m,1}$. Logit-transformed proportion $\psi_{c,t,m,1} = logit(\phi_{c,t,m,1})$ is modelled with a penalized basis-spline (P-spline) regression model:

$$logit(\phi_{c,t,m,1}) = \psi_{c,t,m,1} = \sum_{k=1}^{K} \beta_{c,m,1,k} B_{c,k}(t), \quad (1)$$

where $B_{c,k}(t)$ refers to the kth basis function evaluated in country c, at time t and $\beta_{c,m,1,k}$ is the kth spline coefficient for the public sector supply (s=1) of method m in country c. The basis functions $B(t)$ are constructed using cubic splines. The basis are fitted over the years 1990 to 2025. We segment the timeline using twelve knots, $1 \leq k \leq 10$. This results in a knot point every 3.5 years approximately. The same spacing is used for each country. The knot points occur when the basis function is at its maximum. We align the knot placement of the basis splines with the most recent survey in each country. As the most recent survey year varies by country, the basis splines $B_{c,k}(t)$ also vary by country. An example of the country-specific basis-functions are shown in Figure 2 where the basis are shown as a set of coloured curves plotted over time with the locations of the knots denoted in black dots with dash vertical lines along the x-axis.

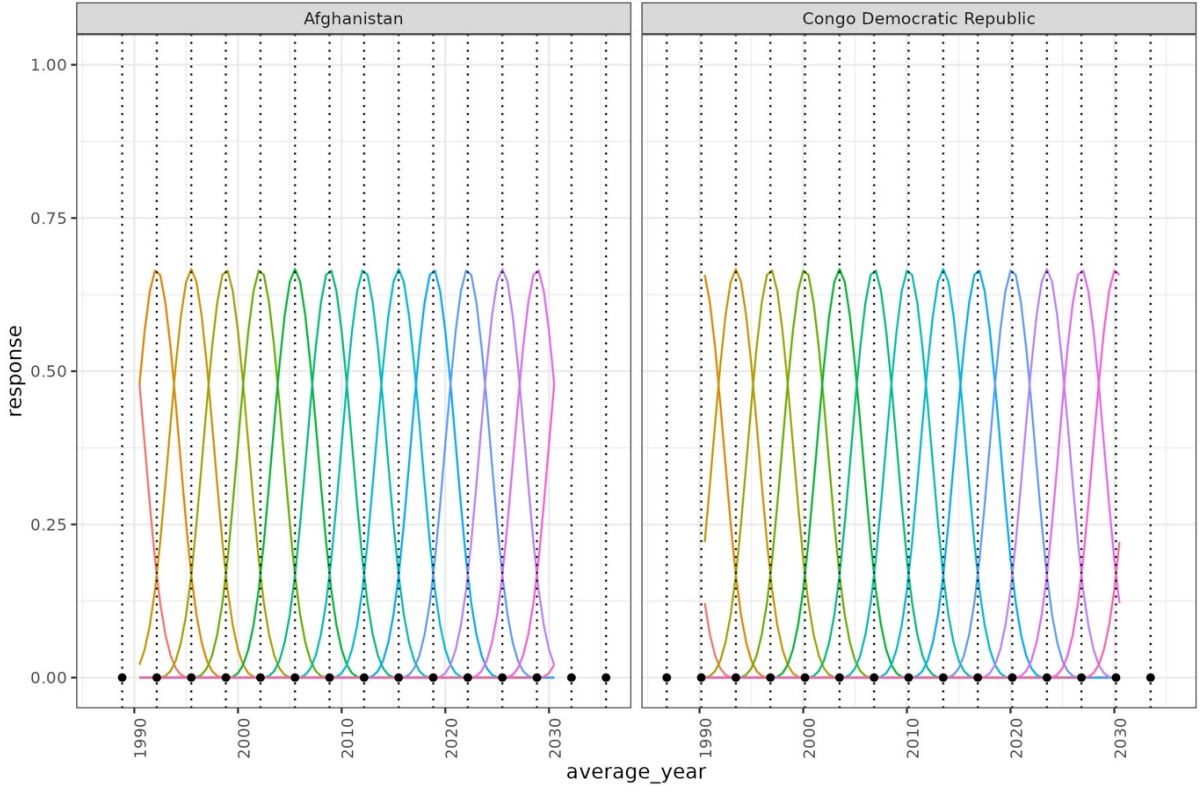

**Figure 2.** The set of basis functions plotted over time that are used to fit Afghanistan and Democratic Republic of Congo (DRC). The most recent survey for Afghanistan occurs in 2015 and the most recent survey in DRC occurs in 2013. Thus, the knot point locations, depicted as black dots with vertical dashed lines differ between the two countries.

Similarly, we model the latent variable, $\psi_{c,t,m,2}$, to capture the logit-transformed ratio of the private commercial medical supply share to the total private sector share. The model is specified as follows:

$$logit\left(\frac{\phi_{c,t,m,2}}{1-\phi_{c,t,m,1}}\right) = \psi_{c,t,m,2} = \sum_{k=1}^{K} \beta_{c,m,2,k} B_{c,k}(t), \quad (2)$$

where $\beta_{c,m,2,k}$ is the kth spline coefficient for the private medical sector supply of method m in country c.

From the latent variable vector, $\boldsymbol{\psi}_{c,t,m}$, it is possible to infer the compositional vector $\boldsymbol{\phi}_{c,t,m}$.

$$\phi_{c,t,m,1} = logit^{-1}(\psi_{c,t,m,1}), \quad (3)$$

$$\phi_{c,t,m,2} = (1 - \phi_{c,t,m,1})logit^{-1}(\psi_{c,t,m,2}), \quad (4)$$

$$\phi_{c,t,m,3} = 1 - (\phi_{c,t,m,1} + \phi_{c,t,m,2}). \quad (5)$$

*3.1.1 Parametrization of the spline regression coefficients*

To estimate the spline coefficients, $\beta_{c,m,s,k}$, we model them indirectly by estimating the penalised first order differences of the spline coefficients, $\boldsymbol{\delta}_{c,m,s}$. The $\boldsymbol{\delta}_{c,m,s}$ vector is of length h where h=K-1 and is defined as

$$\boldsymbol{\delta}_{c,m,s} = (\beta_{c,m,s,2} - \beta_{c,m,s,1}, \beta_{c,m,s,3} - \beta_{c,m,s,2}, \ldots, \beta_{c,m,K} - \beta_{c,m,K-1}). \quad (6)$$

We assume that in country c, method m and sector s, the value of spline coefficient at knot point k*, aligning with the year t* where the most recent survey occurs, is $\alpha_{c,m,s}$. By doing this, we are assuming that the $\alpha_{c,m,s}$ parameter will act as the spline coefficient for the reference spline at k*. We are then able to calculate the spline coefficients from the reference knot (k*) using the estimated $\boldsymbol{\delta}_{c,m,s}$.

$$\beta_{c,m,s,k} = \begin{cases} \alpha_{c,m,s} & k = k^*, \\ \beta_{c,m,s,k+1} - \delta_{c,m,s,k} & k < k^*, \\ \beta_{c,m,s,k-1} + \delta_{c,m,s,k-1} & k > k^*. \end{cases} \quad (7)$$

*Hierarchical estimation of the intercept*

The parameter $\alpha_{c,m,s}$ acts similarly to an intercept term as it forms the baseline level from which forward and backward projections are based off. A change in $\alpha_{c,m,s}$ can lead to a systematic change on the estimated set of supply-share levels for a particular country. $\alpha_{c,m,s}$ is estimated hierarchically to allow parameter estimates to benefit from cross-method and then cross-country information sharing for each sector. The hierarchical distributions are given by:

$$\alpha_{c,m,s} \mid \theta_{r[c],m,s}, \sigma_{\alpha,s}^2 \sim N(\theta_{r[c],m,s}, \sigma_{\alpha,s}^2),$$

$$\theta_{r,m,s} \mid \theta_{w,m,s}, \sigma_{\theta,s}^2 \sim N(\theta_{m,s}^w, \sigma_{\theta,s}^2).$$

Such that $\alpha_{c,m,s}$ is distributed around a sector, method, region-specific mean ($\theta_{r[c],m,s}$) allowing for a cross-method variance ($\sigma_{\alpha,s}^2$) for each sector. The geographic regions, r[c], used to group countries together are the UNSD intermediate world regions[21]. The sector, method, region-specific means are distributed around a sector, method-specific world mean ($\theta_{m,s}^w$) allowing for a cross-regional variance ($\sigma_{\theta,s}^2$) within each sector. We chose this hierarchical setup as the private sector plays an important role in the supply of contraceptive products, such as OC pill and injectables, whereas the public sector tends to provide higher proportions of clinical contraceptive methods, including female sterilisation, IUDs and implants[22,23]. Thus, it made sense for the data to split the prior and hyper-prior mean parameters by method-type to respect these observed differences.

Finally, the overall sector-specific mean is given a vague Normal prior and the cross-method and cross-country variance parameters are given half-Cauchy distributions:

$$\theta_{w,m,s} \sim N(0,100),$$

$$\sigma_{\alpha,s}^2 \sim Cauchy(0,1)T(0,),$$

$$\sigma_{\theta,s}^2 \sim Cauchy(0,1)T(0,).$$

*Specification of the deviation terms δ*

A multivariate normal prior centred on 0 was assigned to the vector of length M of first-order differences of the spline coefficients, $\boldsymbol{\delta}_{c,1:M,s,h}$, for all methods supplied by sector s in country c, at first-order difference h,

$$\boldsymbol{\delta}_{c,1:M,s,h}| \Sigma_{\delta,s} \sim MVN(\mathbf{0}, \Sigma_{\delta,s}),$$

Where,

$$\Sigma_{\delta,s} = \begin{bmatrix} \sigma^2_{\delta_{1,s}} & \hat{\rho}_{1,2,s}\sigma_{\delta_{1,s}}\sigma_{\delta_{2,s}} & \cdots & \hat{\rho}_{1,M,s}\sigma_{\delta_{1,s}}\sigma_{\delta_{M,s}} \\ \hat{\rho}_{2,1,s}\sigma_{\delta_{2,s}}\sigma_{\delta_{1,s}} & \sigma^2_{\delta_{2,s}} & \cdots & \hat{\rho}_{2,M,s}\sigma_{\delta_{2,s}}\sigma_{\delta_{M,s}} \\ \cdot & \cdot & \cdots & \cdot \\ \hat{\rho}_{M,1,s}\sigma_{\delta_{M,s}}\sigma_{\delta_{1,s}} & \cdot & \cdots & \sigma^2_{\delta_{M,s}} \end{bmatrix}. \quad (8)$$

The variance terms $\sigma^2_{\delta_{m,s}}$ for $m = 1,..,M$ and $s = 1, 2$ are method-sector specific smoothness parameters that act as a penalization parameter on the first order differences of the spline coefficients. As $\sigma^2_{\delta_{m,s}}$ tends towards 0, deviations away from the sector, method, country-specific mean go to 0. Within each sector, we assume that the first order differences are not independent across methods. This dependency is captured via the covariance $\hat{\rho}_{i,j,s}\sigma_{\delta_{i,s}}\sigma_{\delta_{j,s}}$ for $s = 1, 2$, $i = 1,..,M$ and $j = 1,..,M$ and where $\hat{\rho}_{i,j,s}$ is the estimated correlation between first-order differences of spline coefficients for method $i$ and method $j$ supplied by sector $s$. Half-Cauchy priors were assigned to the standard deviation terms $\boldsymbol{\sigma}_\delta$, similar those used in the estimation of the intercept.

The correlation terms of the covariance matrix, $\hat{\rho}_{i,j}$, were estimated using a maximum *a posteriori* estimator for the correlation matrix as described in Azose and Raftery, 2018[24]. This approach involves fitting a model where the covariance terms in $\Sigma_{\delta,s}$ are set equal to zero. The resulting estimates are then used to estimate the correlation between methods across time and all countries. Specifically, for sector s, the correlation between method $i$ and method $j$ is calculated as follows,

$$\hat{\rho}_{i,j,s} = \frac{\sum_{c=1}^{C}\sum_{h=1}^{K-1} \tilde{\delta}_{c,m[i],s,h}\tilde{\delta}_{c,m[j],s,h}}{\sqrt{\sum_{c=1}^{C}\sum_{h=1}^{K-1} \tilde{\delta}_{c,m[i],s,h}^2} \sqrt{\sum_{c=1}^{C}\sum_{h=1}^{K-1} \tilde{\delta}_{c,m[j],s,h}^2}}, \quad (9)$$

Where $\tilde{\delta}_{c,m,s,h}$ are the estimated first order differences of the spline coefficients estimated in the zero-covariance model, C represents the total number of countries involved in the study and h represents the number of differences (h=$K_c$-1) between the spline coefficients. The point estimates $\tilde{\delta}_{c,m,s,h}$ are given by the posterior medians of $\delta_{c,m,s,h}$ in the zero-covariance run, after subsetting the period considered to periods with data within a country.

Figure 3 shows the heat map of the estimated correlations between the first-order differences in the spline coefficients for the five contraceptive methods studied for the public sector and ratio of commercial medical sector to total private sector. In general, the public sector has weak to moderate positive correlations across all method combinations (Figure 3). The positive correlations seen in the private sector are stronger than those seen in the public sector. In the private sector, the long-acting and permanent (LAPM) methods female sterilization and injectables have very strong positive relations with both each other and OC pills. In both the public and private sector plots, OC pills and injectables have the strongest relationship (Figure 3.A; 0.54. Figure 3.B; 0.78). In both instances, OC pills have a strong positive correlation with injectables. This implies that the rates of change in the supply of OC pills tend to increase jointly with the rates of change in the supply of injectables across the public and private sectors.

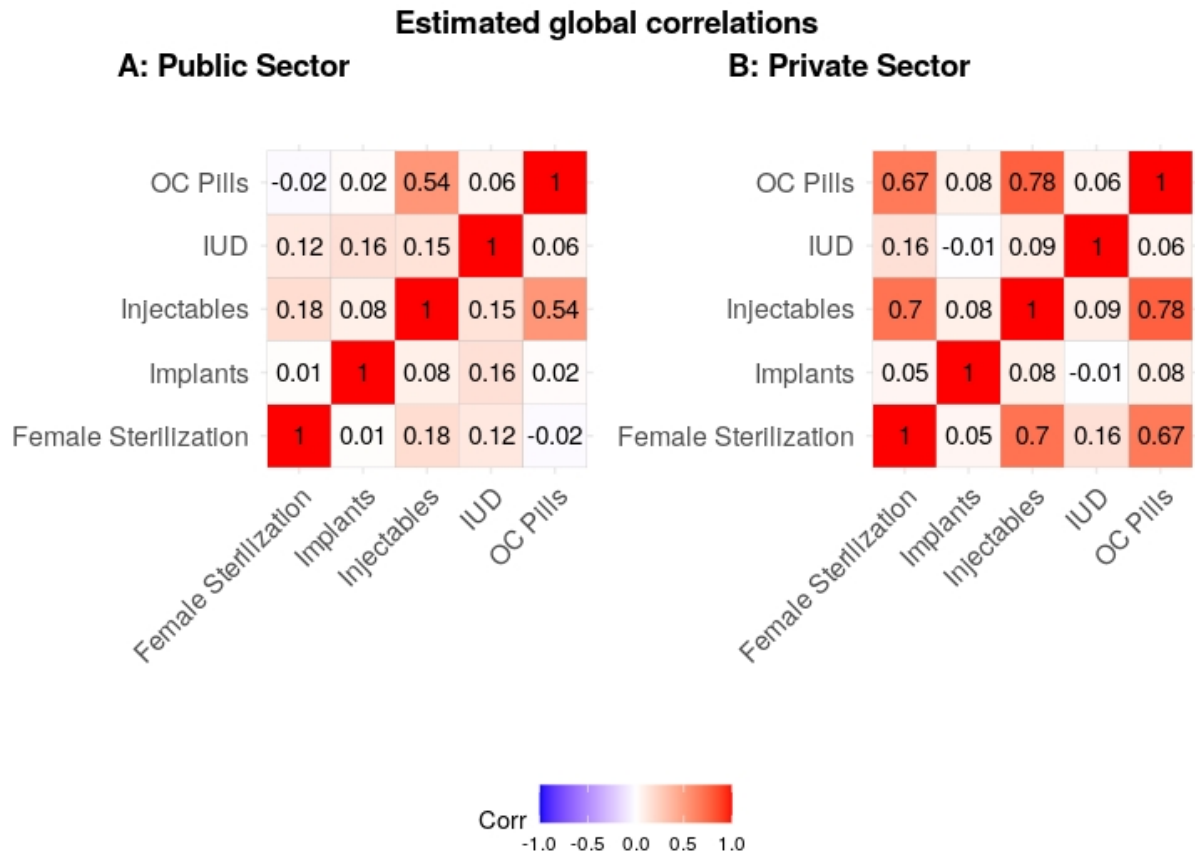

**Figure 3.** A heatmap of the estimated correlations between the changes in contraceptive supply of contraceptive methods (on the logit scale) present in the public (A) and private (B) sectors. The methods include female sterilization, oral contraceptive pills (OC pills) intrauterine devices (IUD), injectables and implants. Correlation is measured on a scale from -1 (strong negative correlation) to +1 (strong positive correlation). The colours represent the strength of the correlation, red indicates negative correlations and blue indicates positive correlations.

*The data model*

The likelihood of the observed data, $y_i$, the observed proportion of modern conceptive method supplied by the public and commercial medical sectors ($s=1$ or $s=2$) for method *m,* at time *t* in country *c,* are modelled using truncated Normal distributions limited to the interval (0,1) such that

$$y_i \mid \phi_{c[i],t[i],m[i],s[i]} \sim \text{N}\big(\phi_{c[i],t[i],m[i],s[i]}, SE_i^2\big)\text{T}(0,1),$$

Where, $\phi_{c[i],t[i],m[i],s[i]}$ is the supply proportion for the country, time-point, method, and sector associated with observation $i$. The variance utilizes the standard error (SE) calculated using the DHS survey microdata associated with observation $y_i$.

*Country Estimates*

We produced estimates and projections of the public and private sector contraceptive supply share from 1990 to 2023 for 5 contraceptive methods in 30 countries involved in FP2030. Results are presented here for a subset of countries. Results for all countries are included in the Appendix (Figures A.1 - A.25). We show results for five countries to illustrate the model's response to varying amounts of data available for each country in the data set. These countries include Afghanistan, Democratic Republic of Congo (DRC), Mozambique, Nepal, and Zimbabwe. Afghanistan has at most a single survey data point for each method and sector (Table 1). The DRC and Mozambique have at most three survey data points for each method and sector (Table 1). Finally, Nepal and Zimbabwe have at least three survey data points for each method and sector (Table 1).

      To begin, we will look at our model estimates in a data-rich setting. Historically, Nepal appears to distribute the pill supply share equally across the public and private sectors (Figure 4A). In contrast to this, Zimbabwe traditionally supplied most OC pills through the public sector (Figure 4B). However, in more recent years, Zimbabwe's private medical and private other sectors have been slowly growing in popularity. For other methods the public sector dominates the supply in both countries, however in Nepal the public sector share of IUDs does appear to be decreasing slowly over time.

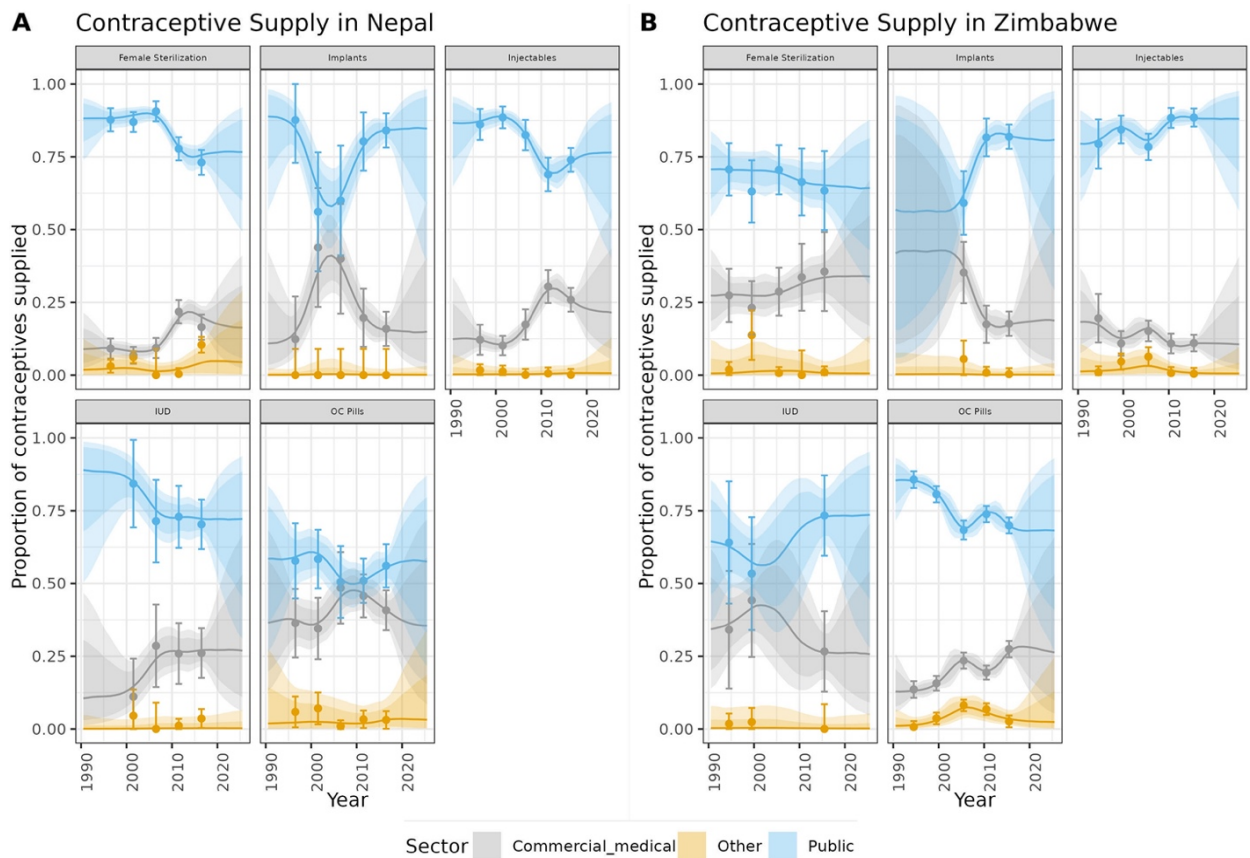

**Figure 4.** The projections for the proportion of modern contraceptives supplied by each sector in Nepal (A) and Zimbabwe (B). The median estimates are shown by the continuous line while the 80% and 95% credible interval is marked by shaded coloured areas. The DHS data point is signified by a point on the graph with error bars displaying the standard error associated with each observation. The sectors are coloured blue for public, grey for commercial medical and gold for other private.

In the DRC, except for the OC pill, the public sector provides the largest share of all contraceptive methods (based on median estimates; Figure 5A). More recently, the proportion of OC pill supplied by the private sector seems to be declining (since ~ 2014). The public sector supply share is increasing over time for female sterilization. Injectables, which in the past projections appeared steady in the public sector supply share, shows an increase in supply from the public sector in more recent years (since ~ 2008). The IUD trend is being impacted by the across-method correlation structure imposed in the model. Changes over time in IUD supply from the public sector has a weak positive correlation with changes in injectables (correlation=0.15) and female sterilization (correlation=0.12) (Figure 3) and these relationships are influencing the estimates of the IUD supply share in the absence of observed data. The behaviour of the OC pill supply share can be explained by the hierarchical structure of the fixed effects term in the model. In the absence of historical data, OC pill estimates begin at the sector, region, method-specific intercept, which is informed by the most recent supply-share level observed, and then the data begins to inform the estimates in more recent years (post-2014 approximately). The extrapolations into the future are steady as the spline coefficients used during estimation are based on the penalised first-order differences which are expected to result in a zero-unit rate of change between coefficients (equation 9).

In Mozambique, the public sector dominates the supply share for all methods (based on median estimates; Figure 5B). The public sector supply share for all methods is mostly steady over time. We see moderate increases in the public supply share of injectables and OC pills between 1995 and 2010 approximately. For implants, the national estimates are tending towards the sector-specific average of most recently observed supply share levels across the larger geographic region in the absence of data.

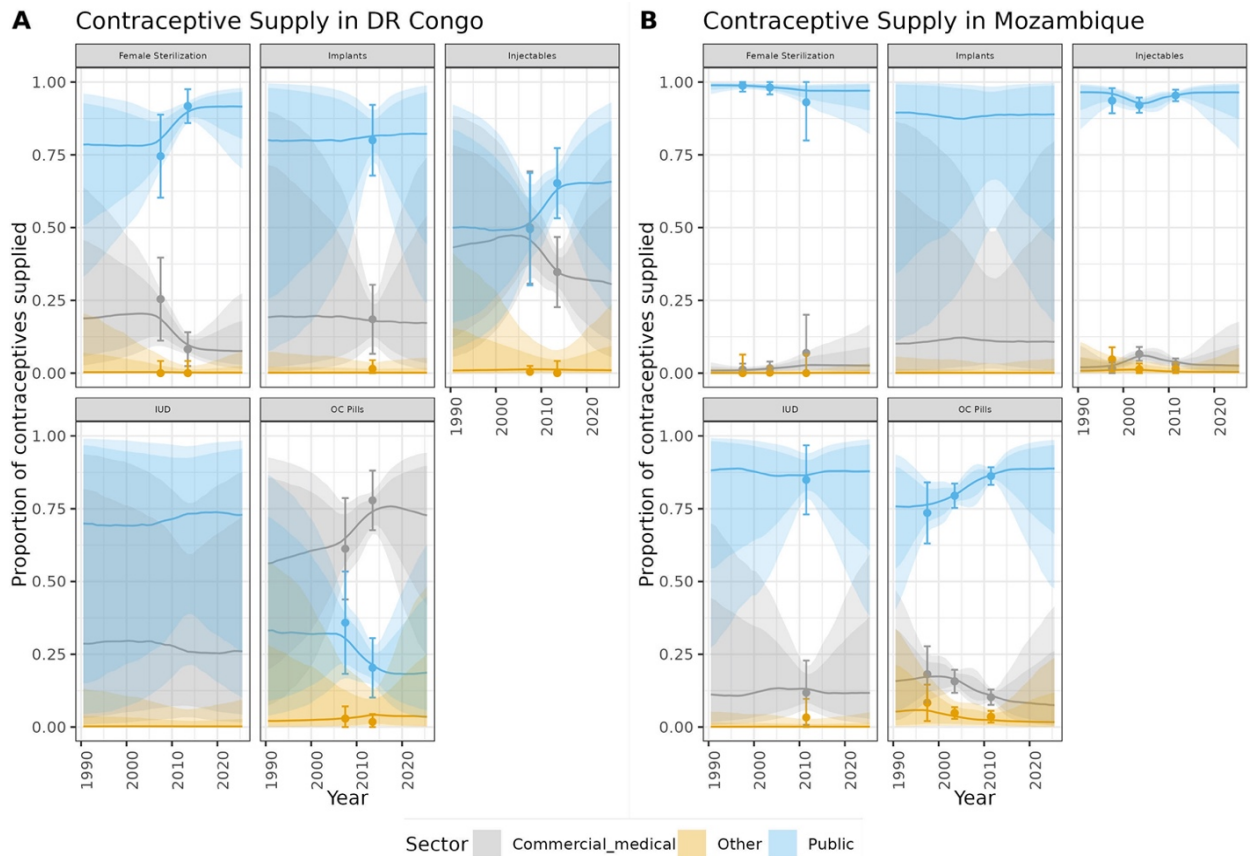

**Figure 5.** The projections for the proportion of modern contraceptives supplied by each sector in Democratic Republic of Congo (DR Congo) (A) and Mozambique (B). The median estimates are shown by the continuous line while the 80% and 95% credible interval is marked by shaded coloured areas. The DHS data point is signified by a point on the graph with error bars displaying the standard error associated with each observation. The sectors are coloured blue for public, grey for commercial medical and gold for other private.

Lastly in Afghanistan (Figure 6), there is only one survey available. For all methods except for the OC pill, the public sector provides the largest supply share over the study period (based on median estimates). The supply share median estimates and uncertainty intervals for female sterilization, injectables, IUD and OC pills are influenced by single data points. In the absence of any data for implants, the estimates are centred on the sector, region, method-specific average of the most recently observed supply share levels.

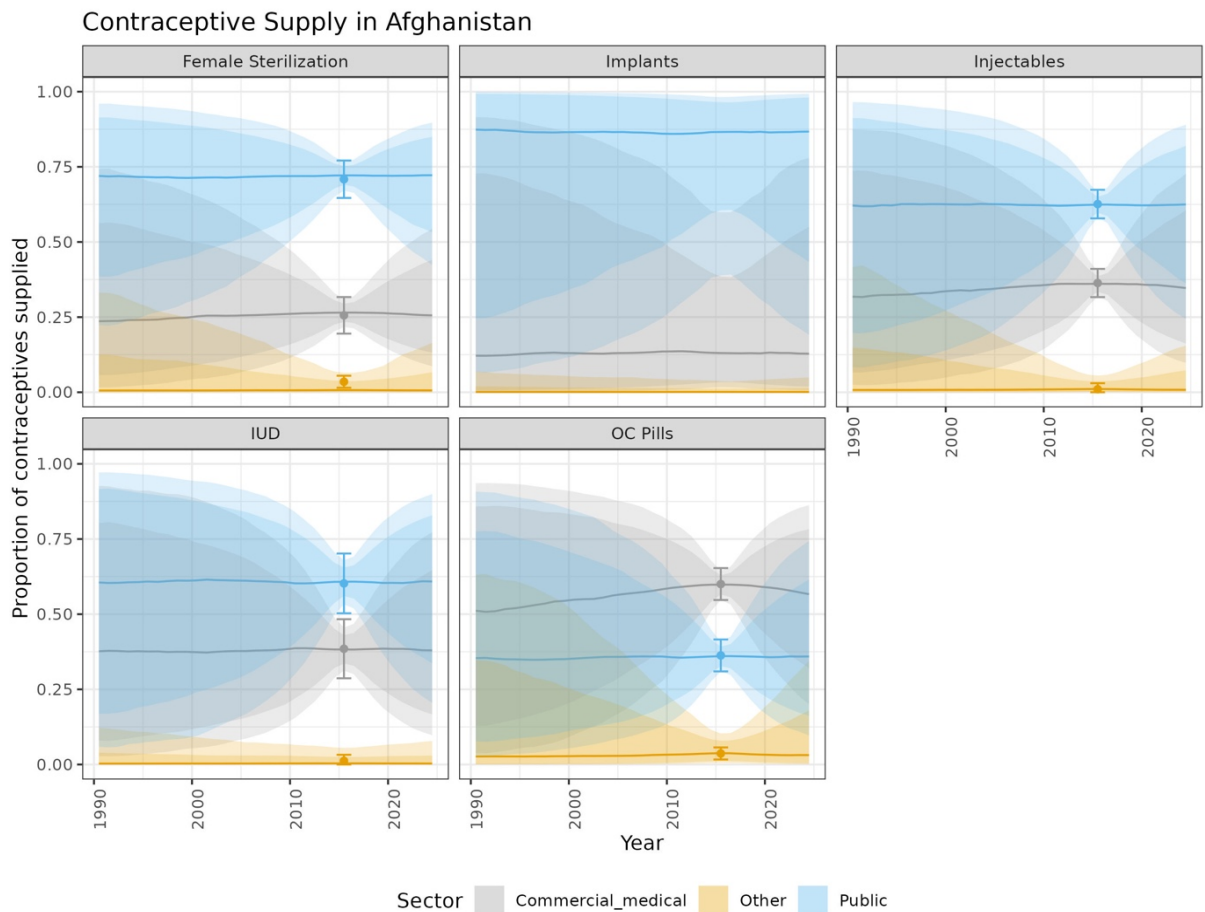

**Figure 6.** The projections for the proportion of modern contraceptives supplied by each sector in Afghanistan. The median estimates are shown by the continuous line while the 80% and 95% credible interval is marked by shaded coloured areas. The DHS data point is signified by a point on the graph with error bars displaying the standard error associated with each observation. The sectors are coloured blue for public, grey for commercial medical and gold for other private.

Table 3 shows a summary of the contraceptive supply share for each method in 2022. Overall, the public sector supplies the highest proportion of each contraceptive method, and the private other sector supplies the smallest proportion. Implant supply is dominated by the public sector (approximately 87%). This may be since they require medical assistance for insertion and therefore require medically trained providers. This makes them more expensive to provide and less likely to be accessed through private means[27]. In contrast, the supply of OC pills is approximately a 45:55 percentage point split between the public and private sectors. OC pills can be supplied to users without medical assistance making them more popular in the private sector[28].

|  | Public | | Private Commercial Medical | | Private Other | |
| --- | --- | --- | --- | --- | --- | --- |
|  | Mean (%) | SD (%) | Mean (%) | SD (%) | Mean (%) | SD (%) |
| Female Sterilization | 82.35 | 11.49 | 16.34 | 10.88 | 1.31 | 1.29 |
| Implants | 83.45 | 12.68 | 16.12 | 12.40 | 0.43 | 0.29 |
| Injectables | 73.57 | 18.82 | 22.65 | 16.48 | 3.78 | 9.65 |
| IUD | 73.02 | 15.13 | 26.23 | 14.77 | 0.75 | 0.41 |
| OC Pills | 45.30 | 21.84 | 42.44 | 20.24 | 12.26 | 10.95 |

**Table 3.** Summary of contraceptive supply share proportions across all countries and in 2022 for each method and sector. The mean estimate for the percentage point supplied and the associated standard deviation (SD) are listed.

**Discussion**

In this paper we proposed and validated a Bayesian hierarchical model to estimate the country and method specific changes in the contraceptive supply share over time with (sometimes) limited amounts of DHS survey data. The modelling framework uses penalised splines to capture the evolution of the contraceptive method supply from the public and private sectors while imposing a correlation structure to capture correlations between changes in the supply share. The penalised splines provide a flexible fit to the data without overfitting.

The model structure imposes a hierarchy such that the expected sector, country, method-specific supply shares are informed based on the geographical relationships of the countries to promote information sharing across countries where smaller amounts of data are present. The model will produce estimates within the period of available survey data as well as projections beyond the most recent data point.

The model was used to estimate and project the contraceptive supply share from the public, private commercial medical and private other sectors for 30 focus countries of the FP2030 initiative, where relevant data was available. Case study examples illustrated the strength and flexibility of the modelling approach in its ability to capture the evolving nature of the contraceptive supply share over time. The hierarchical setup of the model and the imposed cross-method correlation structure allows the model to produce informed estimates (with uncertainty) even in cases where very limited data are available. Based on a series of validation measures, the proposed model is well calibrated and outperforms various model alternatives (see Supplementary Appendix).

An application of our model will be in the calculation of Estimated Modern Use (EMU), a proxy for modern contraceptive prevalence derived from service statistics and frequently used in the Family Planning Estimation Model[7,29]. A full description of how to calculate EMUs can be found in the Appendix, section 'Calculating Estimated Modern Use'. Service statistics data include the number of family planning commodities distributed to clients, commodities distributed to health facilities, family planning facility visits, or family planning facility users. Frequently, the raw data reported in a given country does not include the private sector contribution. As such, the raw data may not be fully representative of the country's whole contraceptive market. Currently, in order to address this issue when calculating EMUs[6], the most recent DHS survey estimate of the country's contraceptive supply share is used to provide a breakdown of the contraceptive market and scale up the raw data accordingly. We instead propose to use our model-based estimates in this important supply-adjustment step of the EMU calculation and to propagate the posterior uncertainty for the contraceptive supply share of sector s, method m, at time t in country c into the EMU calculations.

Another potential application of these model estimates could be used in evaluating the security of contraceptive method supply chains. It is well-established that a total-market approach (TMA) is key to both the longevity and sustainability of the family planning market and in achieving equity among family planning users[28]. Understanding and quantifying the supply shares of the private and public sectors in the contraceptive market contributes to a TMA approach to supply chain management[30]. Presently, the latest DHS survey estimates are used to provide the proportion of modern contraceptives supplied by each sector. However, this relies on out-of-date information in relation the private sector supply in the absence recent surveys. The model estimates presented here can provide an essential measure of the private sector's role in modern contraceptive method supplies at a national level. Using the model estimates, it is possible to predict and reflect on overall contraceptive market supply trends. This feeds into indicator 1, "Generating intelligence", and indicator 6, "Accountability and transparency in health markets", of the TMA framework which will help to enable evidence-based strategies for resource management within the family planning market and improve the sustainability of future family planning initiatives[31].

The estimates and projections presented here provide important and relevant information on the contraceptive supply share over time, at a national level. The data model allows for the incorporation of survey sampling errors to be included in the modelling process. The use of penalised-splines allows for data driven, flexible model-based estimates and the correlation structure imposed in the estimation of the splines allows the model to draw strength and information from the relationships between supply-shares of different contraceptive methods. The Bayesian hierarchical framework allows for information sharing across methods and regions, which is of particular use for countries where data are limited. This consideration of the public/private sector supply of contraceptive methods within countries is vital for informing family planning programs that seek to improve access to and increase use of modern family planning methods[32]. The private sector can play an important role in the sustainability of family planning markets and therefore the ability, via model-based estimates, to evaluate where countries currently stand in terms of private sector contributions to the contraceptive method supply can help to inform needs for private sector expansion and support the growth of contraceptive use[32].

**Data availability**

The code and data used to create these results can be found at *mcmsupply*.

**Acknowledgements**

This publication has emanated from research conducted with the financial support of Science Foundation Ireland under Grant number 18/CRT/6049. This work was supported, in whole or in part, by the Bill & Melinda Gates Foundation [INV-008441]. Under the grant conditions of the Foundation, a Creative Commons Attribution 4.0 Generic License has already been assigned to the Author Accepted Manuscript version that might arise from this submission.

**Disclosure of interest**

The authors report there are no competing interests to declare.

# Appendix

**Calculating the Standard Errors using the DHS Microdata**

Using the '*svyciprop*' function from the '*survey*' package in R[1], the proportions of each method supplied the public and privates sectors for the specific surveys in the countries listed above were calculated along with their associated standard errors. The '*survey*' package uses the Taylor series linearisation method to approximate the standard error of the calculated proportions[2,3].

| Measure | Range<br>(% over all methods) | Median SE size<br>(% over all methods) | Largest mean SE<br>(Method, %) | | Smallest mean SE<br>(Method, %) | |
|---|---|---|---|---|---|---|
| Result | 0.015, 22.6 | 2.72 | IUD | 4.71 | Injectables | 2.52 |

**Table A.1.** Summary table for the calculated standard errors of the data observations.

The calculated standard errors range from 0.015 to 22.6 percentage points. The median standard error size across all method is 2.72 percentage points. On average, they tend to be largest for IUDs where the mean standard error size is 4.71 percentage points and smallest in injectables where the mean standard error size is 2.52 percentage points.

*Creating the Estimates*

We used R and JAGS (Just Another Gibbs Sampler) to fit the model. JAGS uses Markov Chain Monte Carlo (MCMC) algorithm to produce model estimates for Bayesian Hierarchical models[25]. To evaluate the JAGS output we used *'rjags'*, an R package that offers cross-platform support from JAGS to the R interface[26]. The results were a set of trajectories for the proportion of contraceptive *m* supplied by each sector over time for each country included in the study. The median of these results was taken to be the model's point estimate. 80% credible intervals were calculated using the 10th and 90th percentiles while 95% credible intervals were calculated using the 2.5th and 97.5th percentiles from the posterior distribution for each estimate.

*Model Validation*

*Out-of-sample validation*

The out-of-sample validation involves training the model on all but the most recent observation for countries that had two or more DHS data points. The withheld data then becomes a test set which we judge our models' projections against. The training set had 350 observations while the test set had 112 observations. We used the model to produce estimates and projections with 95% prediction intervals for all countries contained within the training set for the years 1990 to 2025. We then compared the resulting projections to the test set of most recent observations to check model performance.

*Errors and coverage*

We calculate sector specific error terms, $e_{j,s}$, to describe the difference between the observed data point $j$ in sector s, $y_{j,s}$, and the median estimate from the posterior predictive distribution, $\hat{y}_{j,s}$.

$$e_{j,s} = y_{j,s} - \hat{y}_{j,s}.$$

We evaluated the results of the validation using different measures of accuracy and prediction interval calibration. To evaluate the accuracy of our model, we considered the root mean square error (RMSE) for each sector's set of estimates.

$$RMSE_s = \sqrt{\frac{\sum_{j=1}^{N_s}(e_{j,s})^2}{N_s}}. \quad (A.1)$$

In equation A.1, $N_s$ is the number of observations in the sector *s*, and $e_{j,s}$ is the error for observation j in sector s which is described above. The RMSE can be interpreted as the average error observed across all countries, time points and methods in the test set.

We also evaluated the mean error and the median absolute errors produced by the model. The mean error is the average difference between the observed proportion and true proportion estimated by the model and is an effective measurement of bias within the model. When the mean error is positive, this indicates systematic under-prediction by the model and conversely, a negative mean error indicates that the model is over-estimating the observed data. Median absolute error is the 50$^{th}$ percentile of absolute differences between the observed proportion and true proportion estimated by the model. Median absolute error captures the overall variation within the model estimates.

$$Mean\ Error_s = \frac{\sum_{j=1}^{N_s} e_{j,s}}{N_s}, \qquad (A.2)$$

$$Median\ Absolute\ Error_s = Median\left(\sum_{j=1}^{N_s} |e_{j,s}|\right). \qquad (A.3)$$

In equation A.2 and A.3, $N_s$ and $e_{j,s}$ are as described above in equation A.1.

Coverage assumes that if our model is correctly calibrated, then for each sector the model should be able to capture the test set of out-of-sample observations with 95% accuracy, where the remaining 5% of incorrectly estimated observations are approximately evenly distributed above and below the estimated 95% prediction interval. To examine the bias of our models estimates, we examined the location of the incorrectly estimated test set observations. We consider the proportion of test observations located above and below the estimated prediction intervals. By examining the breakdown of locations, we are evaluating the tendency of the model to under- or over-estimate the test set. If a higher proportion of observations are located below the prediction intervals, this indicates that the model is tending to over-estimate the test set. Similarly, if a higher proportion of the incorrectly estimated observations are located above the prediction intervals, the model is tending to under-estimate the test set.

*Justification of model complexity*

To evaluate whether the complexity of this model is required, we compare the complex model validation results to those of a linear regression model and a P-spline regression model without any cross-method correlation. Details of these models and their set-up can be found in the below section on 'Model Comparison'.

**Results**

*Out-of-sample model validation*

The model described in this paper has been evaluated using various out-of-sample model validation measures to evaluate its effectiveness at estimating the method supply shares at a national level while also considering the prediction intervals it uses to produce these estimates. It is performing reasonably well considering the complex nature of the data. It has an overall coverage of approximately 93%. The public sector and private commercial medical sector estimates have similar validation results. This model outperforms simpler modelling alternatives, described in the appendix.

The results for the out of sample validation are found in Table A.2. The target coverage is 95%. The model is reasonably well calibrated to the data with both the public sector and commercial medical sector having 91% coverage. The private other sector has 96% coverage of the test set. Across all three sectors, the overall coverage is approximately 93%.

The mean error of all sectors is positive and less than 1 percentage point. The median absolute error of the private commercial medical sector was the largest at 6.5 percentage points while the median absolute error of the private other sector was the smallest at approximately 1.1 percentage points.

| Sector | Mean error (%) | MAE (%) | RMSE (%) | Median Prediction Interval Width (%) | Breakdown of location for the incorrectly estimated observations | | Coverage (Target 95%) |
|---|---|---|---|---|---|---|---|
| | | | | | Above the prediction interval (%) | Below the prediction interval (%) | |
| Private Commercial Medical | 0.2 | 6.5 | 12.2 | 39.1 | 5.4 | 3.6 | 91.1 |
| Private Other | 0.4 | 1.1 | 6.4 | 13.8 | 1.8 | 1.8 | 96.4 |
| Public | 0.9 | 6.4 | 12.4 | 39.2 | 5.4 | 3.6 | 91.1 |

**Table A.2.** Leave-one-out validation results for the test set using the model described in the main paper. MAE is median absolute error. RMSE is root mean square error. Coverage is the proportion of the test set observations that are captured within the 95% prediction interval produced by the model. We consider the median PI width and evaluate the location of the incorrectly estimated leave-one-out validation test set observations.

For the private commercial medical sector, we see that the RMSE of this sector was 12.2% which indicates that the average error across all countries and methods was approximately 12 percentage points (Table A.2.) (For comparison with alternative models, see Table A.3 & A.4). The private other sector has smallest the RMSE at approximately 6 percentage points. This is in line with the magnitude of the proportions found in this sector. In general, the private other makes up a small proportion of the overall sector mix. Finally for the public sector, the RMSE also indicates an average error of approximately 12 percentage points. The RMSE of this sector is marginally larger than the other two sectors (12.4%, Table A.2.) but notably, the proportions supplied by the public sector are often the highest across all methods and countries.

The public and private commercial medical sectors have similar median prediction interval widths. The public sector is marginally larger but they both cover approximately a 39-percentage point range. The smallest median prediction interval width was produced by the private other sector covering approximately 14-percentage points (Table A.2.).

When examining the location of the test observations not captured within the prediction intervals, we expect to see approximately 2.5% of the test observations below and above the prediction interval. An even distribution of the incorrectly estimated observations would indicate unbiased estimation of the test set. Both the private commercial medical and public sectors see larger proportions outside the PI than expected, with both sectors having coverage of 91.1% respectively (Table A.2.). This is slightly below the expected 95% mark. Both sectors have a larger proportion of incorrectly estimated test set observations above the PI. This indicates that the incorrectly estimated test observations are higher than the model estimated. The private other sector has an even distribution of incorrectly estimated observations above and below the prediction interval boundaries.

**Model Comparison**

To justify the complexity of our model, we compared it against the performance of two control models – a linear model and a penalized spline model with zero correlation.

*The Linear Model*

To begin, we evaluated the necessity of using a flexible penalized spline approach by comparing our model to an inflexible linear model. The linear model setting is very similar to the model set up described in the main paper, the key difference between the main paper model and this linear model is in the derivation of the latent variables, $\zeta_{c,t,m}$ and $\psi_{c,t,m}$. In the linear model, we describe the latent variables with:

$$logit(\phi_{c,t,m,1}) = \zeta_{c,t,m} = \alpha_{c,m,1} + \delta_{c,m,s} * t \quad (A.4)$$

$$logit\left(\frac{\phi_{c,t,m,2}}{1-\phi_{c,t,m,1}}\right) = \psi_{c,t,m} = \alpha_{c,m,2} + \delta_{c,m,s} * t \quad (A.5)$$

Where,

$\alpha_{c,m,s}$ is a sector, method, country-specific intercept as described by the section 'Hierarchical estimation of the intercept' of the main paper.

$\delta_{c,m,s}$ is a sector, method, country-specific slope term.

The slope term $\delta_{c,m,s}$ is described using the multivariate normal distribution to utilize the linear correlations between observed rates of change in the supply share for each method on the logit scale for the public sector and ratio of commercial medical sector to total private sector.

$$\delta_{c,m,s} \sim MVN(0, \Sigma_{\delta,s})$$

The $\delta_{c,m,s}$ is centered on 0 and the covariance matrix $\Sigma_{\delta,s}$ is of size $m$ x $m$. $\Sigma_{\delta,s}$ is set up according to Equation 9 of the main paper. The components of a compositional vector $\phi_{c,t,m} = (\phi_{c,t,m,1}, \phi_{c,t,m,2}, \phi_{c,t,m,3})$ denoting the proportion $\phi_{c,t,m,s}$ supplied by the public sector ($s$ =1), the private commercial medical sector ($s$=2) and the other private sector ($s$=3) of modern contraceptive method $m$, at time $t$, in country $c$, are derived as described in Equations 3-5 of the main paper.

*The Zero-Covariance Model*

To evaluate the impact of using correlated method-specific covariance terms in the variance-covariance matrix of first-order differences of the spline coefficients, $\delta_{c,m,s}$, we considered a model with zero covariance.

In this zero-covariance model, the model set up is as described in the main paper (Equations 1-8) with the only difference occurring in the calculation of the variance-covariance matrix $\Sigma_{\delta,s}$. As before, the first-order differences of the spline coefficients, $\delta_{c,m,s}$ are described using a multivariate normal prior centred on 0.

$$\delta_{c,m,s} \sim MVN(0, \Sigma_{\delta,s})$$

However, this model the cross-method correlations are set to 0, resulting in zero-covariance on the off-diagonal of the matrix. Thus,

$$\Sigma_{\delta,s} = \begin{bmatrix} \sigma^2_{\delta_{s,1}} & 0 & \ldots & 0 \\ 0 & \sigma^2_{\delta_{s,2}} & \ldots & 0 \\ . & . & \ldots & . \\ 0 & . & \ldots & \sigma^2_{\delta_{s,M}} \end{bmatrix} \quad (A.6)$$

For both the linear and zero-covariance models, we carried out leave-one-out validation using the same training and test data as described in the main paper. The training set was made of 350 training set observations and 112 test set observations. The test set contained the most-recent survey observation for the proportion supplied by the public sector, the private commercial medical sector, and the other private sector of modern contraceptive method m, in country c. We evaluated the models using the same validation measures as those described above in the 'Model Validation' section.

| Sector | Coverage (Target 95%) | Mean error (%) | MAE (%) | RMSE (%) | Median PI Width (%) | Proportion of observations located above/below the estimated PI boundary | |
|---|---|---|---|---|---|---|---|
| | | | | | | Above the PI | Above the PI |
| Private Commercial Medical | 70.5 | 0.322 | 8.5 | 16.5 | 22.9 | 16.99 | 16.99 |
| Private Other | 88.4 | -1.45 | 0.7 | 10.3 | 21.3 | 2.68 | 2.68 |
| Public | 68.8 | 1.54 | 6.75 | 15.6 | 20.1 | 20.50 | 20.50 |

**Table A.3.** Leave-one-out validation results for the test set using linear model. Coverage is the proportion of the test set observations that are captured within the 95% prediction interval (PI) produced by the model. The MAE is the median absolute error. RMSE is root mean square error.

| Sector | Coverage (Target 95%) | Mean error (%) | MAE (%) | RMSE (%) | Median PI Width (%) | Proportion of observations located above/below the estimated PI boundary | |
|---|---|---|---|---|---|---|---|
| | | | | | | Above the PI | Below the PI |
| Private Commercial Medical | 91.96 | 0.13 | 6.25 | 12.29 | 38.92 | 5.36 | 2.68 |
| Private Other | 96.43 | -0.29 | 1.02 | 6.48 | 13.45 | 1.79 | 1.79 |
| Public | 91.07 | 0.84 | 6.66 | 12.36 | 39.90 | 5.36 | 3.57 |

**Table A.4.** Leave-one-out validation results for the test set using zero-covariance model. Coverage is the proportion of the test set observations that are captured within the 95% prediction interval (PI) produced by the model. The MAE is the median absolute error. RMSE is root mean square error.

Coverage measures how well a model is calibrated to predict future observations. In this instance, it is the proportion of test observations that are captured in the 95% prediction interval during leave-one-out cross validation. The full model and zero-covariance model perform similarly as both are reasonably well calibrated for all three sectors. For private commercial medical estimates, the full model had approximately 91% coverage (Table A.2.) while the zero-covariance had slightly higher at approximately 92% (Table A.4). Both models have 91% coverage of the public sector (Table A.2., Table A.4.). Both the zero-covariance and full models have 96% coverage in the private other sector. The coverage of the linear model (Table A.3.) is lower than both the zero-covariance and full models across all sectors. None of the three sectors are optimally fitted in the linear model. As the full model and zero-covariance models have better coverage than the linear model, it justifies the use of complex penalised splines to produce model estimates.

To evaluate the bias and variance produced by the three models, we consider the mean errors, median absolute errors (MAE) and root mean square errors (RMSE). Mean error is a measure of model bias (Equation A.2.). If the mean error for a sector is positive, this indicates systematic under-prediction of test set estimates, conversely a negative mean error indicates potential bias for over-predicting the test set estimates. The median absolute error (Equation A.3.) highlights the variation within the model. A large MAE would indicate that there exists a large difference between the test set observations and the corresponding model estimates. The root mean square error (RMSE) can be interpreted as the average error produced by the model (Equation A.1.).

Both the full model (Table A.2.) and zero-covariance model (Table A.4.) produce positive mean errors for the public sector test set (0.9% and 0.8% respectively). This indicates there is a bias for over-predicting these observations in both models, but as these mean errors are less than 1% they are small. Similarly, the linear model (Table A.3.) has a positive mean error of 1.54%. This mean error is larger than that produced by either spline model. This indicates that the linear model has a larger lack of fit and bias when predicting the public sector estimates. The full model and zero-covariance model also produce positive mean errors for the private commercial medical sector, indicating there is a tendency to overestimate this sector too. The private other sector is evenly estimated in the full model (1.8%, Table A.2.) and underestimated in both the zero-covariance model (-0.29%, Table A.4.) and the linear model (-1.45%, Table A.3.). The linear model produces the largest mean errors for the public and private other sectors across all three models. This indicates it is the least well-calibrated to the data as it has the largest biases. The zero-covariance and full models are performing similarly in terms of mean error.

The median absolute errors (MAE) of the full model (Table A.2.) are like those produced by the zero-covariance model (Table A.4.) for all three sectors. The MAE for the private sector is higher in the linear model than those of the spline models (Table A.3.). The MAE of the private commercial medical sector linear model is 8.50% compared to the full model at 6.5% and the zero-covariance model at 6.3%. The MAE of both the full model and the 0-covariance model are very similar. This indicates that they are both similarly approximating the test set proportions across all three sectors.

The RMSE of the linear model is the largest in all three sectors across the three models (Table A.3.). This indicates that the linear model does not capture the proportions well as the average errors produced are large. The RMSE of the zero-covariance (Table A.4) and full model (Table A.2.) indicate that the average error produced by these models are similar to one another.

When considering this result with the median prediction interval widths, we see that the full model has similar sized prediction interval widths as the zero-covariance model for the public sector (Table A.2., Table A.4). The prediction widths of the full model are smaller in the public and private commercial medical sectors than those produced by the zero-covariance model. This allows the full model to be more effective in capturing the test-set observations while not producing model estimates that are uncertain and uninformative. The linear model had the narrowest PI interval widths of all three models for the public (20.1%) and private commercial medical (22.9%) sectors (Table A.3.). However, considering this metric in tangent with the coverage and errors suggests that the linear model is not a good fit for the data. The PI are too tight and are missing the observations.

Lastly, when considering the location of the incorrectly estimated test set observations. The full and zero-covariance models tend to under-estimate the public sector and the commercial medical sectors (Table A.2.) (Table A.4). The linear model also tends to under-estimate these observations (Table A.3), as a larger proportion of incorrectly estimated public sector observations are found above the PI boundary (20.5%), rather than below it (10.7%). In the private commercial medical sector, the linear model (Table A.3.) tends to under-estimate the test set observations. The incorrectly estimated observations of the private other sector are non-biased in both the full model (Table A.2.) and zero-covariance models (Table A.4.) as there is an equal proportion of observations above and below the PI. The linear model tends to over-estimate the private other sector the most of all three models (Table A.3.; above = 23.1%, below = 76.9%).

Overall, the full model described in the main paper is the most suitable model to describe this data. It captures the complex shape and relationships without overfitting it or missing the shape. It incorporates information regarding the correlations between the rates of change across the contraceptive methods. The average errors and measures of bias produced by the model are similar to the zero-covariance models. The 95% prediction intervals in the full model are marginally tighter than the 0-covariance model while still performing well in the coverage of the test set. The strength of the full model is seen in the absence of data for a particular contraceptive method, where model estimates can still be informed by the behaviour of related methods to produce realistic estimates.

**Country Estimates**

The following section includes the model estimates for the countries not included in the main paper. In each graph, the median estimates are shown by the continuous line while the 80% and 95% credible interval are marked by shaded coloured areas. The DHS data point is signified by a point on the graph with error bars displaying the standard error associated with each observation. Each sector is represented by a different colour, the public sector is blue, the private commercial medical sector is grey while the private other sector is gold.

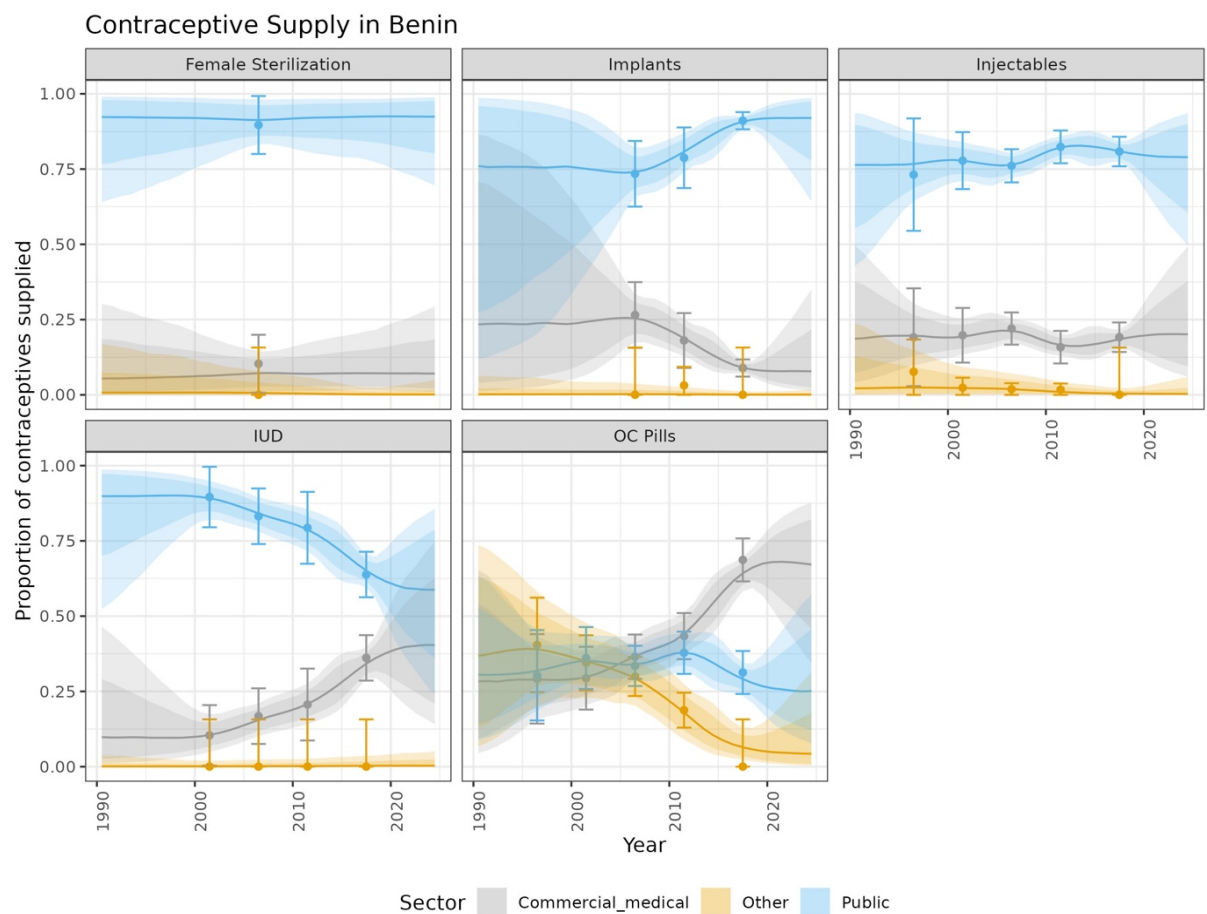

**Figure A.1.** The projections for the proportion of modern contraceptives supplied by each sector in Benin. The median estimates are shown by the continuous line while the 80% and 95% credible interval is marked by shaded coloured areas. The DHS data point is signified by a point on the graph with error bars displaying the standard error associated with each observation. The sectors are coloured blue for public, grey for commercial medical and gold for other private.

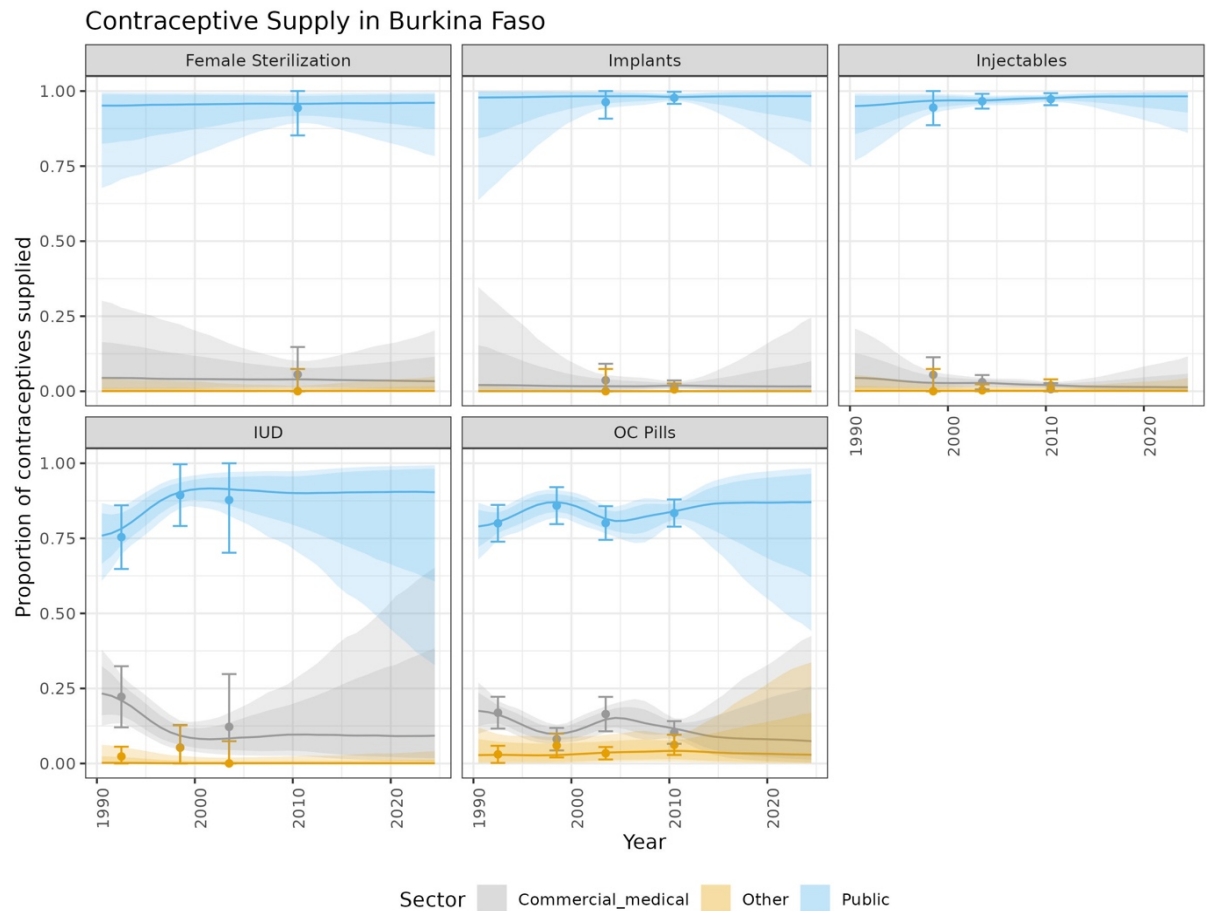

**Figure A.2.** The projections for the proportion of modern contraceptives supplied by each sector in Burkina Faso. The median estimates are shown by the continuous line while the 80% and 95% credible interval is marked by shaded coloured areas. The DHS data point is signified by a point on the graph with error bars displaying the standard error associated with each observation. The sectors are coloured blue for public, grey for commercial medical and gold for other private.

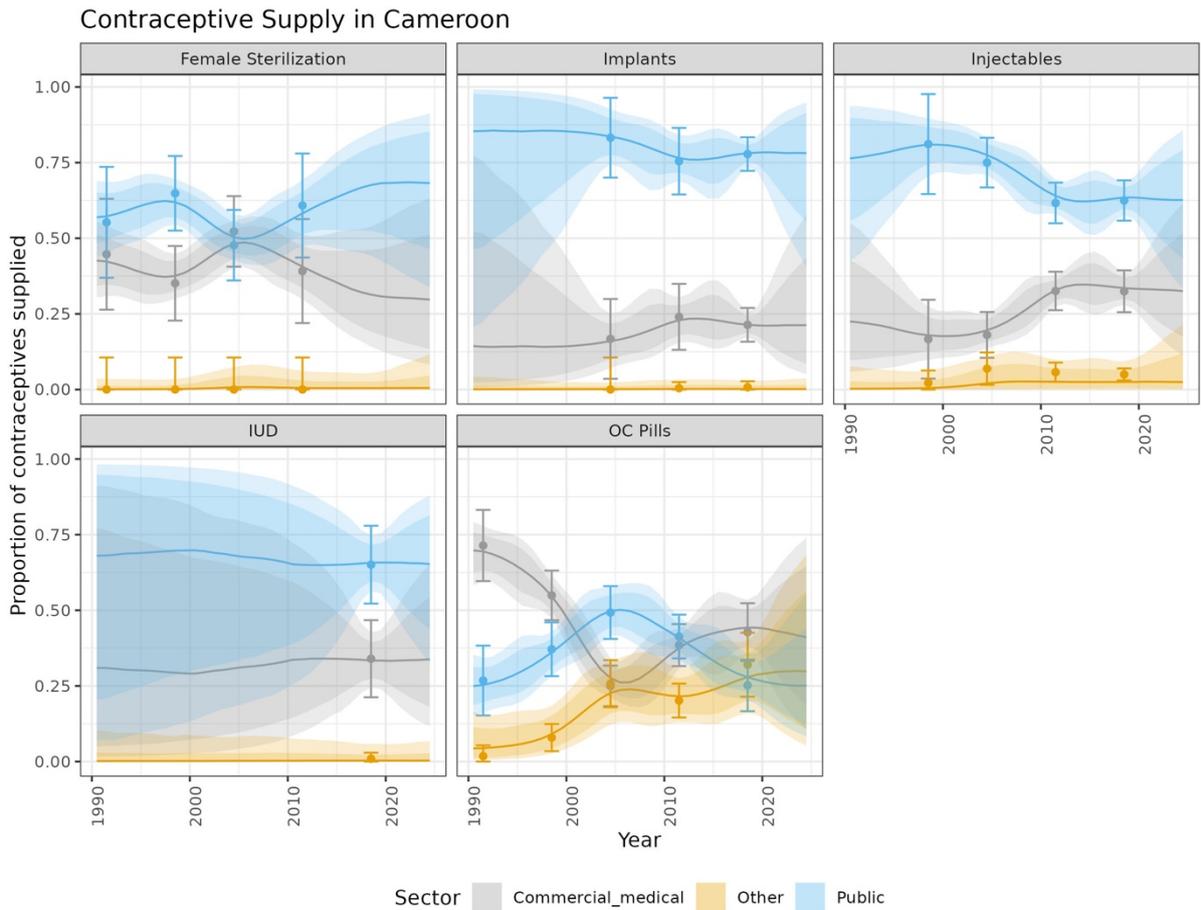

**Figure A.3.** The projections for the proportion of modern contraceptives supplied by each sector in Cameroon. The median estimates are shown by the continuous line while the 80% and 95% credible interval is marked by shaded coloured areas. The DHS data point is signified by a point on the graph with error bars displaying the standard error associated with each observation. The sectors are coloured blue for public, grey for commercial medical and gold for other private.

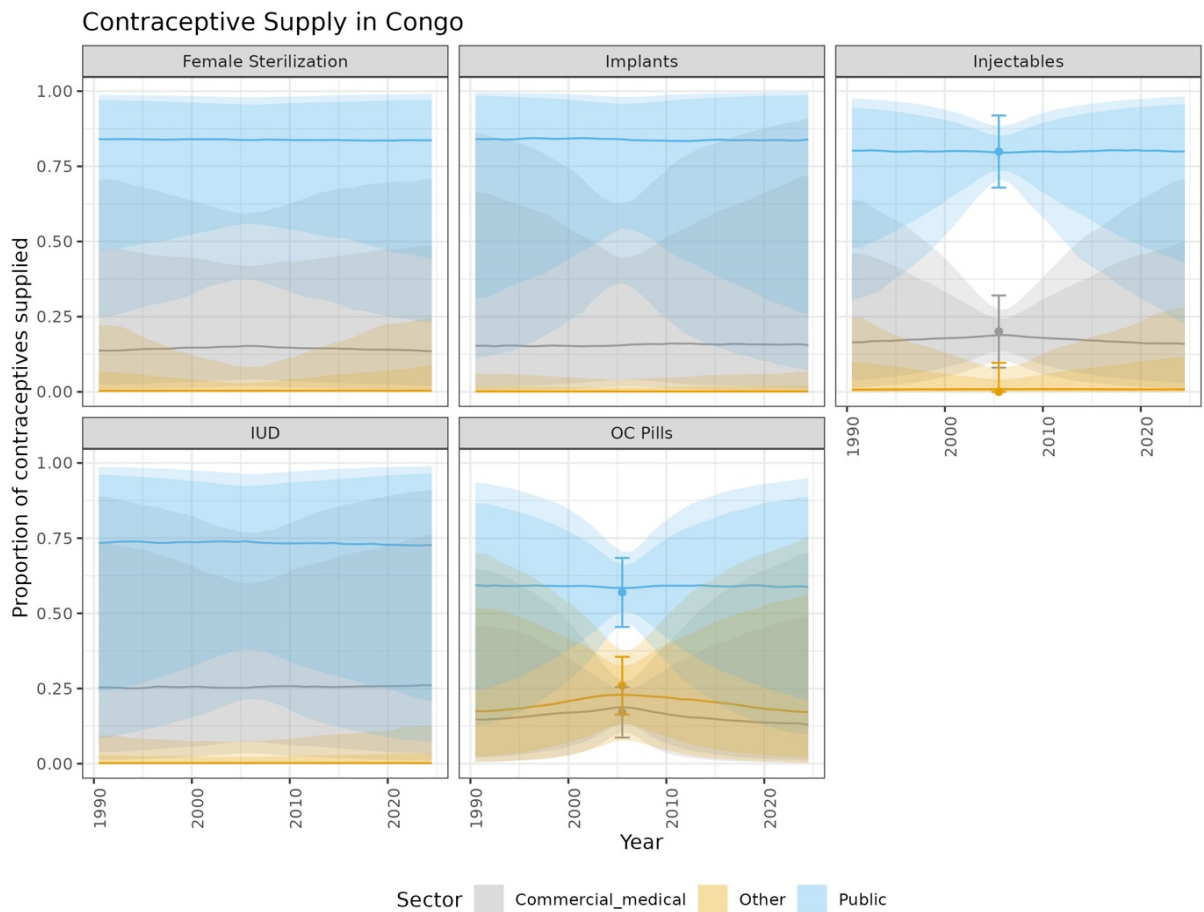

**Figure A.4.** The projections for the proportion of modern contraceptives supplied by each sector in Congo. The median estimates are shown by the continuous line while the 80% and 95% credible interval is marked by shaded coloured areas. The DHS data point is signified by a point on the graph with error bars displaying the standard error associated with each observation. The sectors are coloured blue for public, grey for commercial medical and gold for other private.

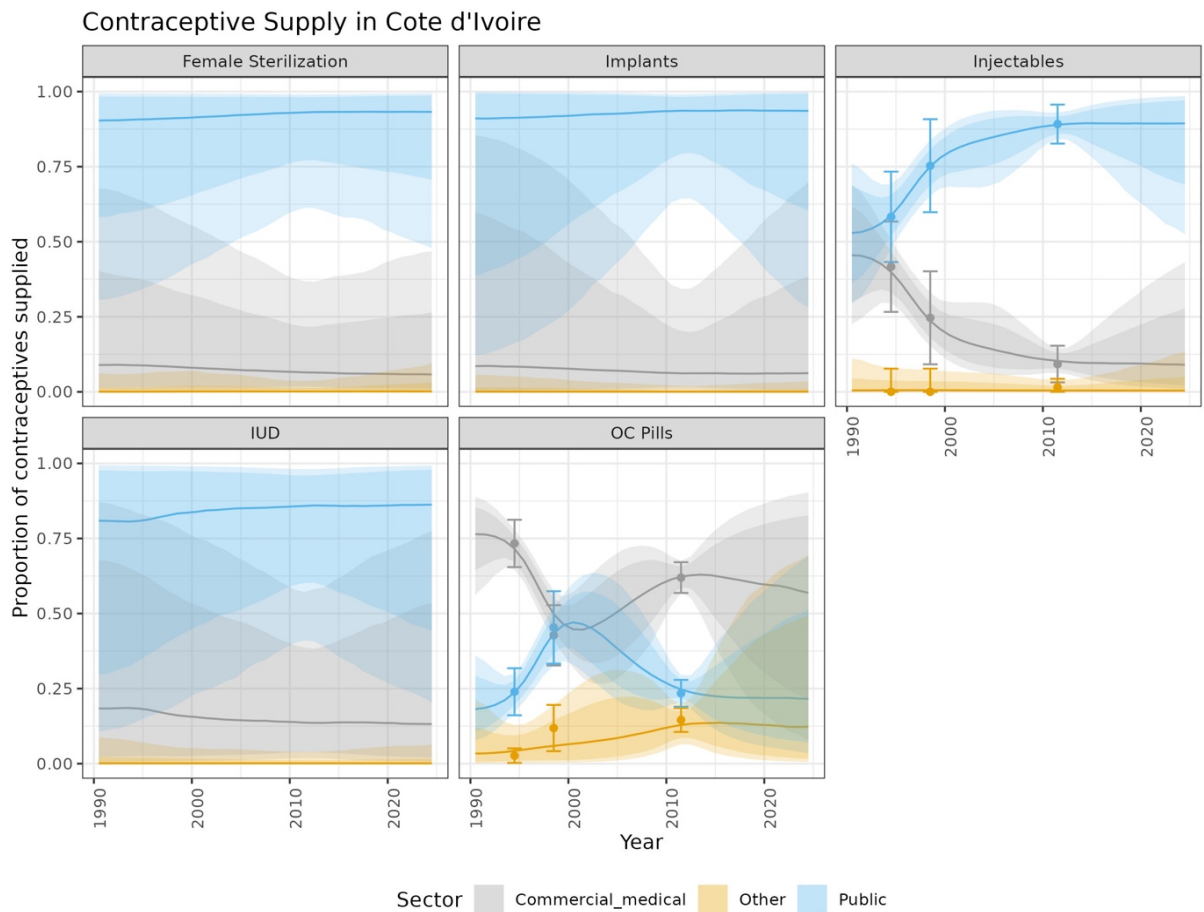

**Figure A.6.** The projections for the proportion of modern contraceptives supplied by each sector in Cote d'Ivoire. The median estimates are shown by the continuous line while the 80% and 95% credible interval is marked by shaded coloured areas. The DHS data point is signified by a point on the graph with error bars displaying the standard error associated with each observation. The sectors are coloured blue for public, grey for commercial medical and gold for other private.

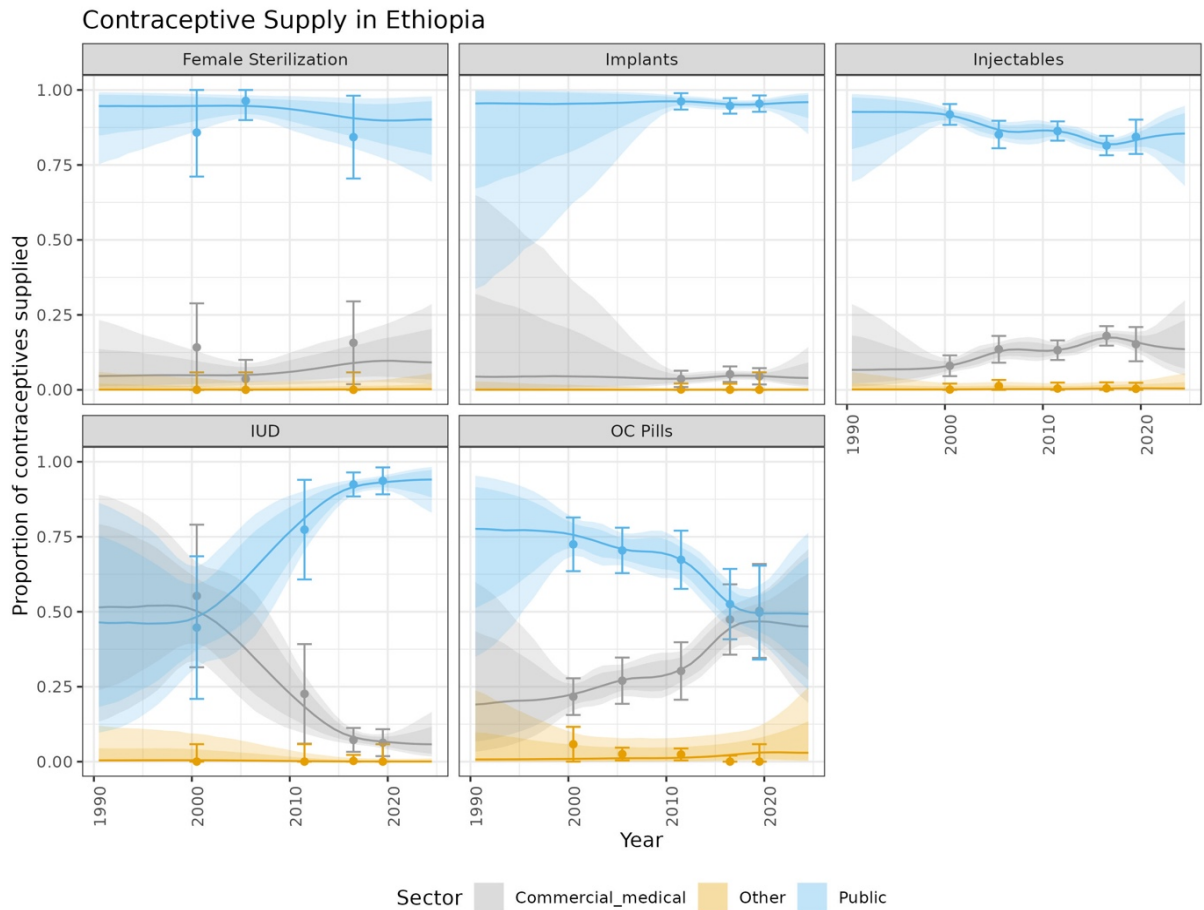

**Figure A.6.** The projections for the proportion of modern contraceptives supplied by each sector in Ethiopia. The median estimates are shown by the continuous line while the 80% and 95% credible interval is marked by shaded coloured areas. The DHS data point is signified by a point on the graph with error bars displaying the standard error associated with each observation. The sectors are coloured blue for public, grey for commercial medical and gold for other private.

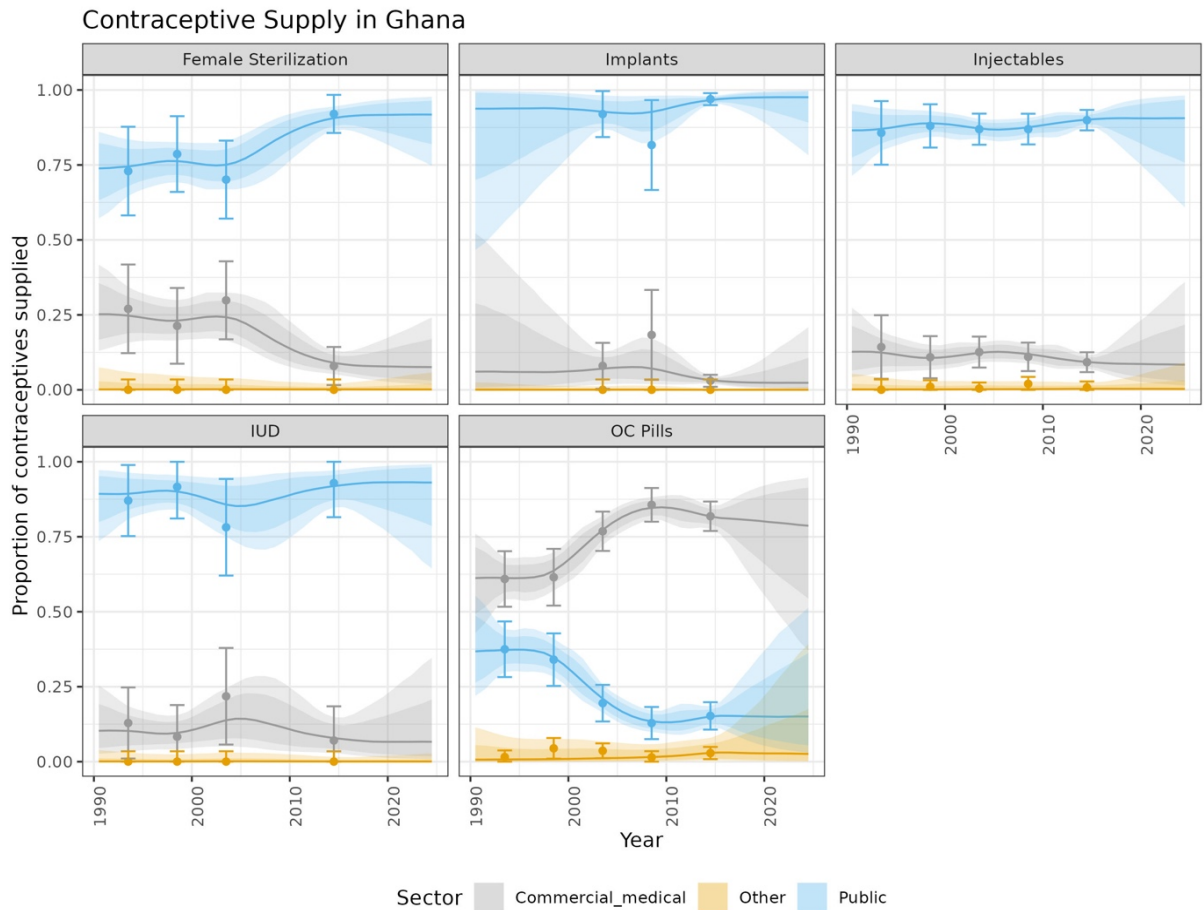

**Figure A.7.** The projections for the proportion of modern contraceptives supplied by each sector in Ghana. The median estimates are shown by the continuous line while the 80% and 95% credible interval is marked by shaded coloured areas. The DHS data point is signified by a point on the graph with error bars displaying the standard error associated with each observation. The sectors are coloured blue for public, grey for commercial medical and gold for other private.

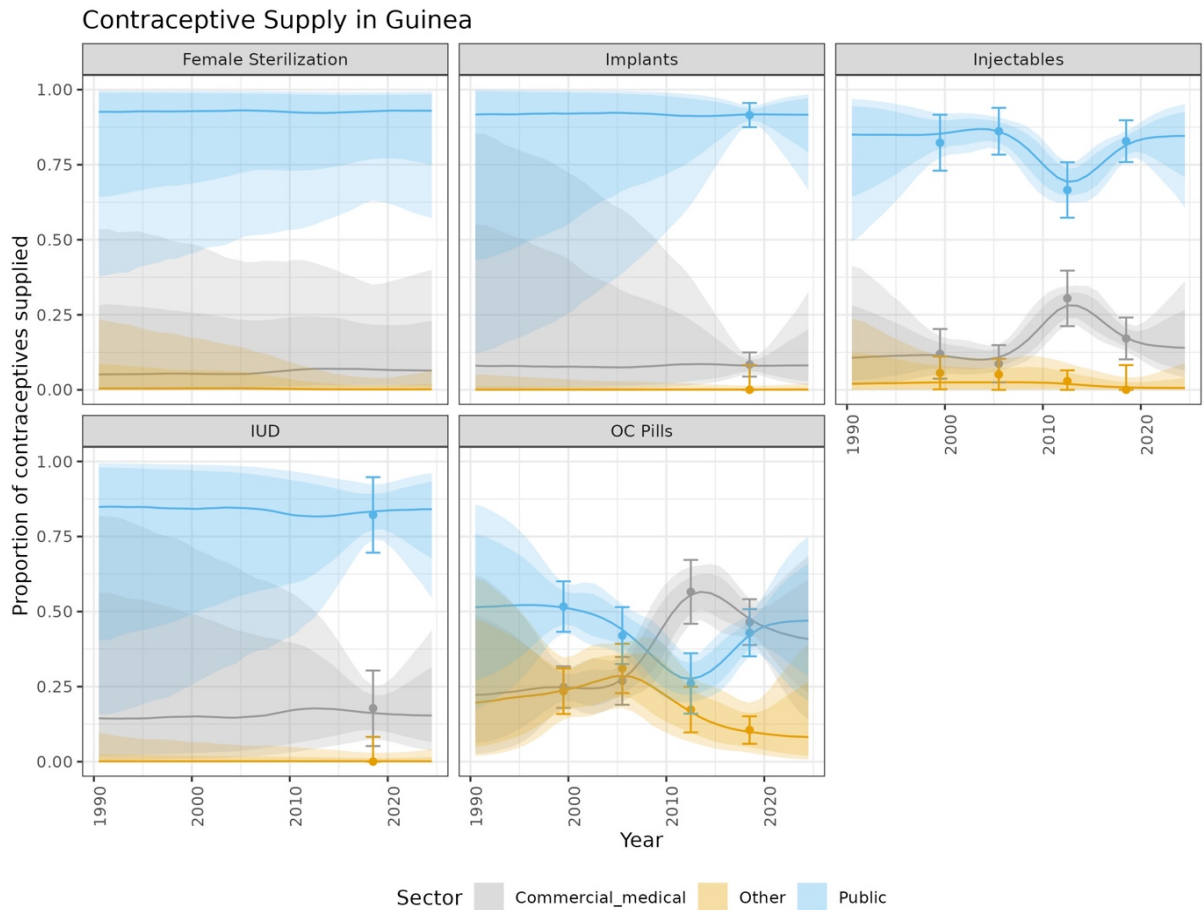

**Figure A.8.** The projections for the proportion of modern contraceptives supplied by each sector in Guinea. The median estimates are shown by the continuous line while the 80% and 95% credible interval is marked by shaded coloured areas. The DHS data point is signified by a point on the graph with error bars displaying the standard error associated with each observation. The sectors are coloured blue for public, grey for commercial medical and gold for other private.

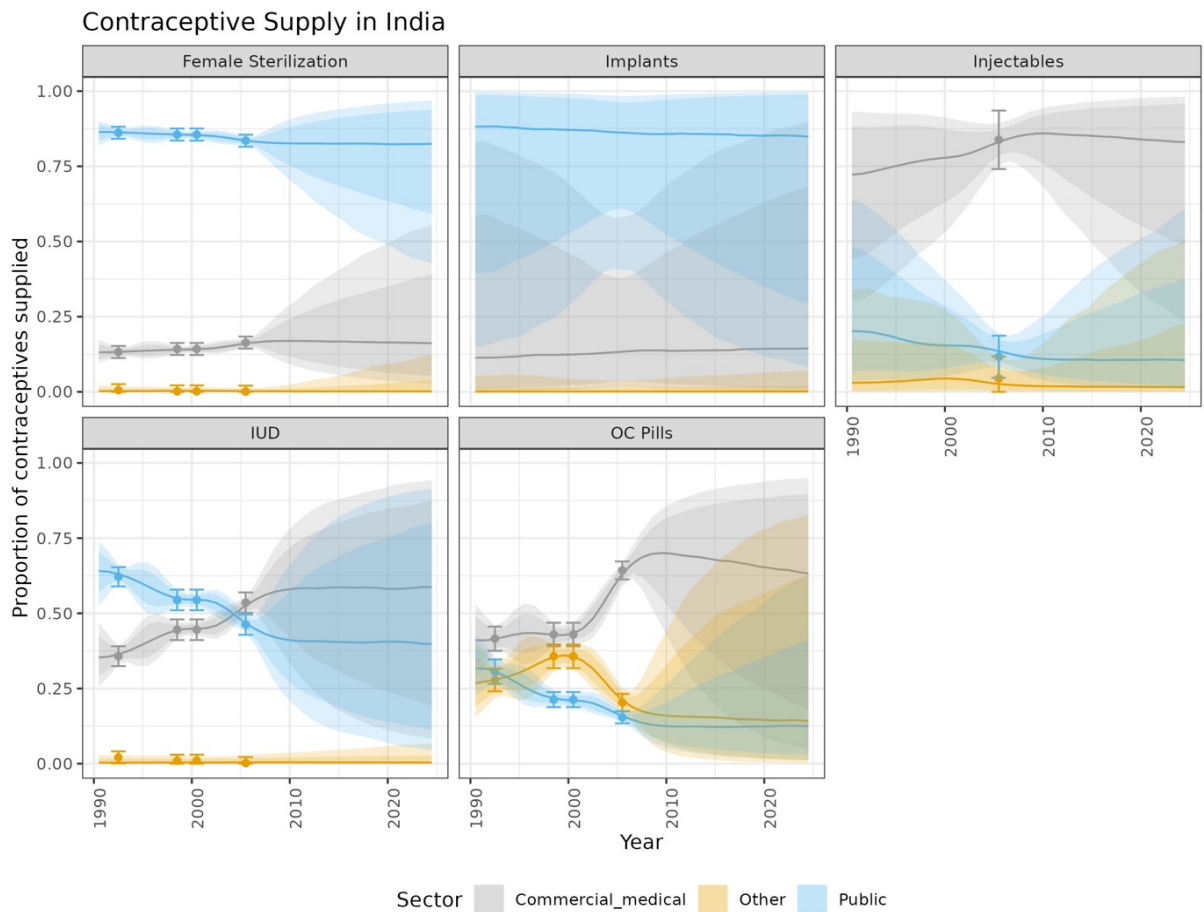

**Figure A.9.** The projections for the proportion of modern contraceptives supplied by each sector in India. The median estimates are shown by the continuous line while the 80% and 95% credible interval is marked by shaded coloured areas. The DHS data point is signified by a point on the graph with error bars displaying the standard error associated with each observation. The sectors are coloured blue for public, grey for commercial medical and gold for other private.

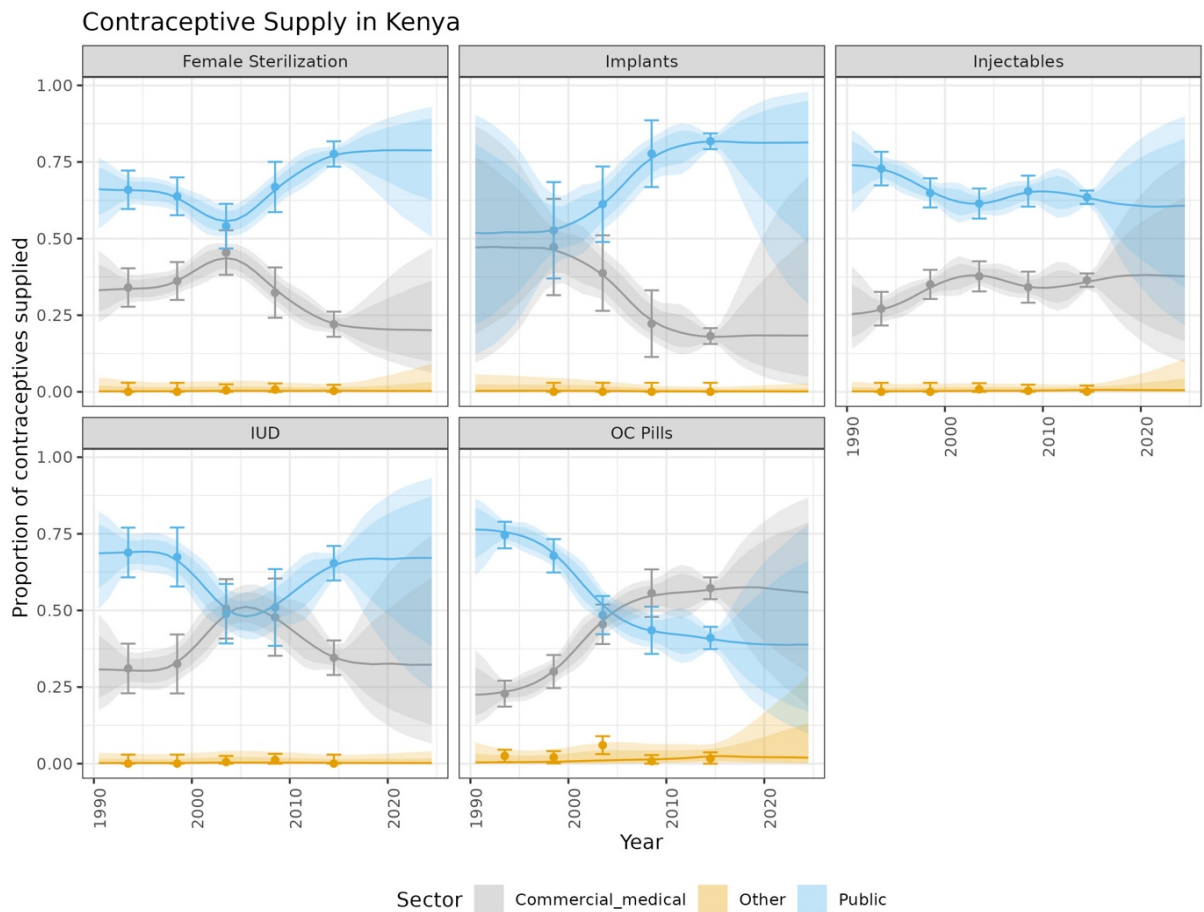

**Figure A.10.** The projections for the proportion of modern contraceptives supplied by each sector in Kenya. The median estimates are shown by the continuous line while the 80% and 95% credible interval is marked by shaded coloured areas. The DHS data point is signified by a point on the graph with error bars displaying the standard error associated with each observation. The sectors are coloured blue for public, grey for commercial medical and gold for other private.

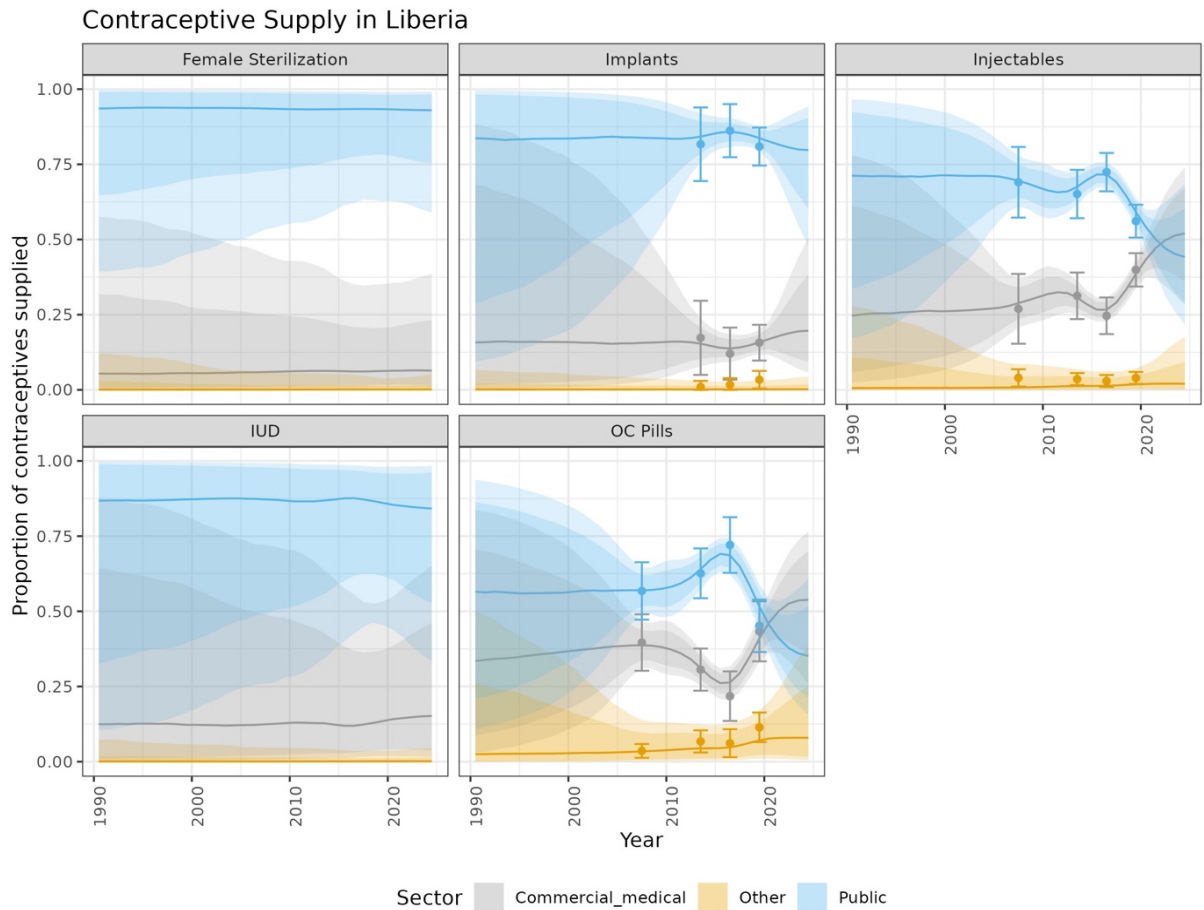

**Figure A.11.** The projections for the proportion of modern contraceptives supplied by each sector in Liberia. The median estimates are shown by the continuous line while the 80% and 95% credible interval is marked by shaded coloured areas. The DHS data point is signified by a point on the graph with error bars displaying the standard error associated with each observation. The sectors are coloured blue for public, grey for commercial medical and gold for other private.

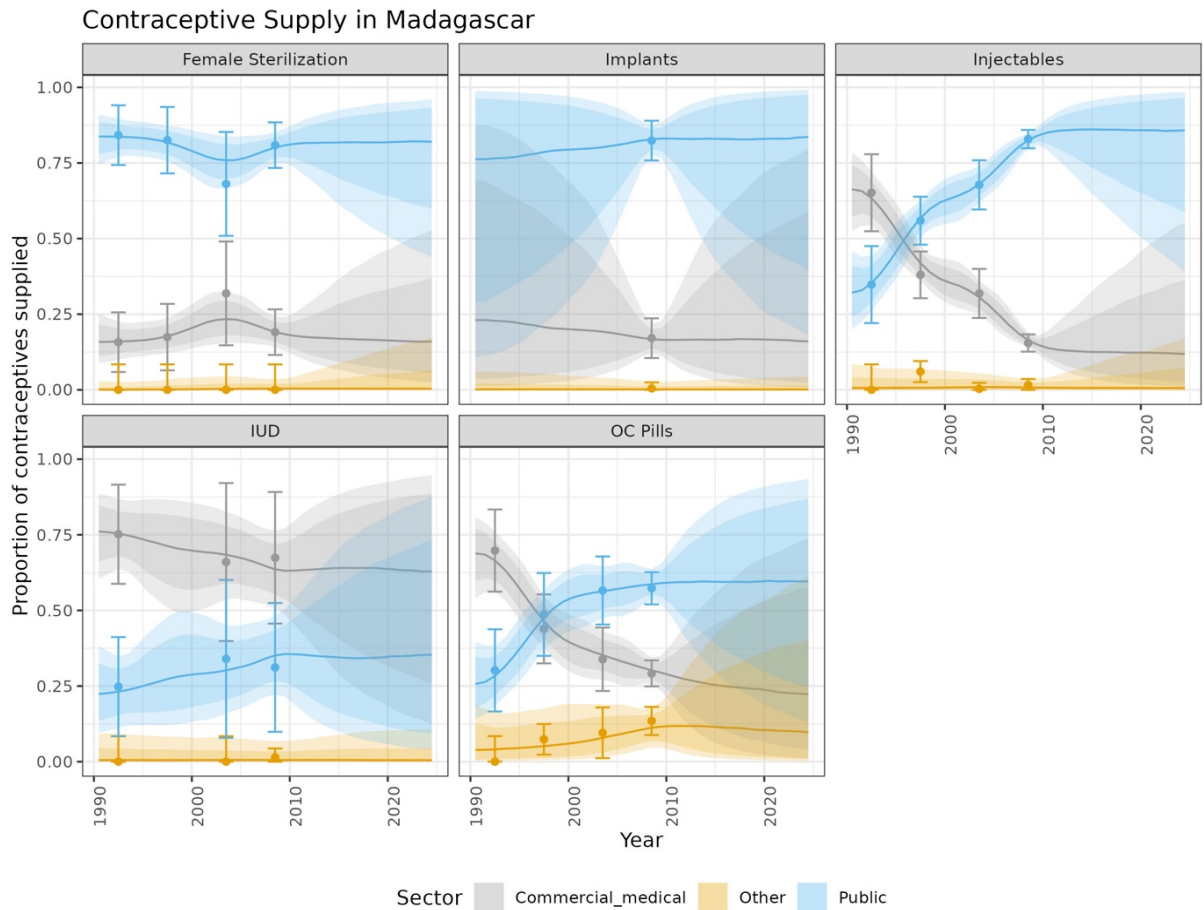

**Figure A.12.** The projections for the proportion of modern contraceptives supplied by each sector in Madagascar. The median estimates are shown by the continuous line while the 80% and 95% credible interval is marked by shaded coloured areas. The DHS data point is signified by a point on the graph with error bars displaying the standard error associated with each observation. The sectors are coloured blue for public, grey for commercial medical and gold for other private.

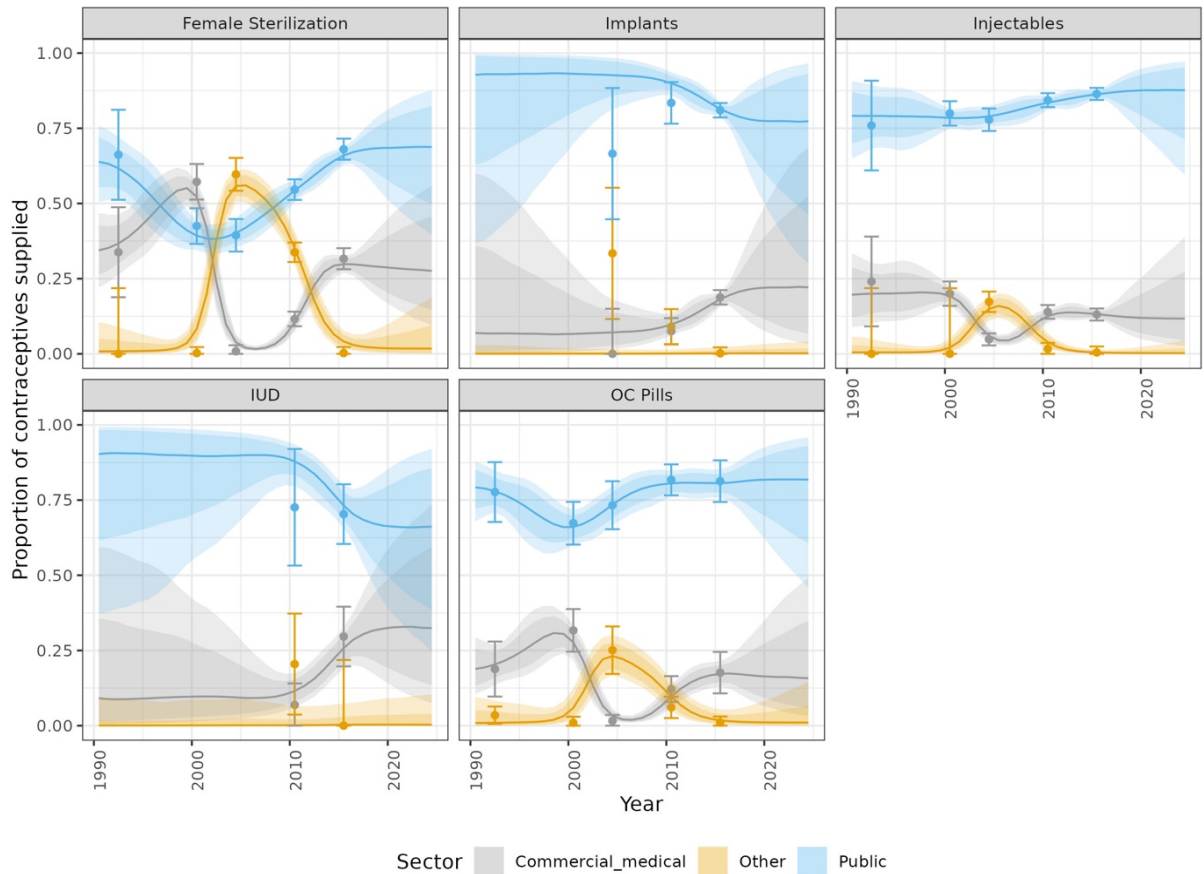

**Figure A.13.** The projections for the proportion of modern contraceptives supplied by each sector in Malawi. The median estimates are shown by the continuous line while the 80% and 95% credible interval is marked by shaded coloured areas. The DHS data point is signified by a point on the graph with error bars displaying the standard error associated with each observation. The sectors are coloured blue for public, grey for commercial medical and gold for other private.

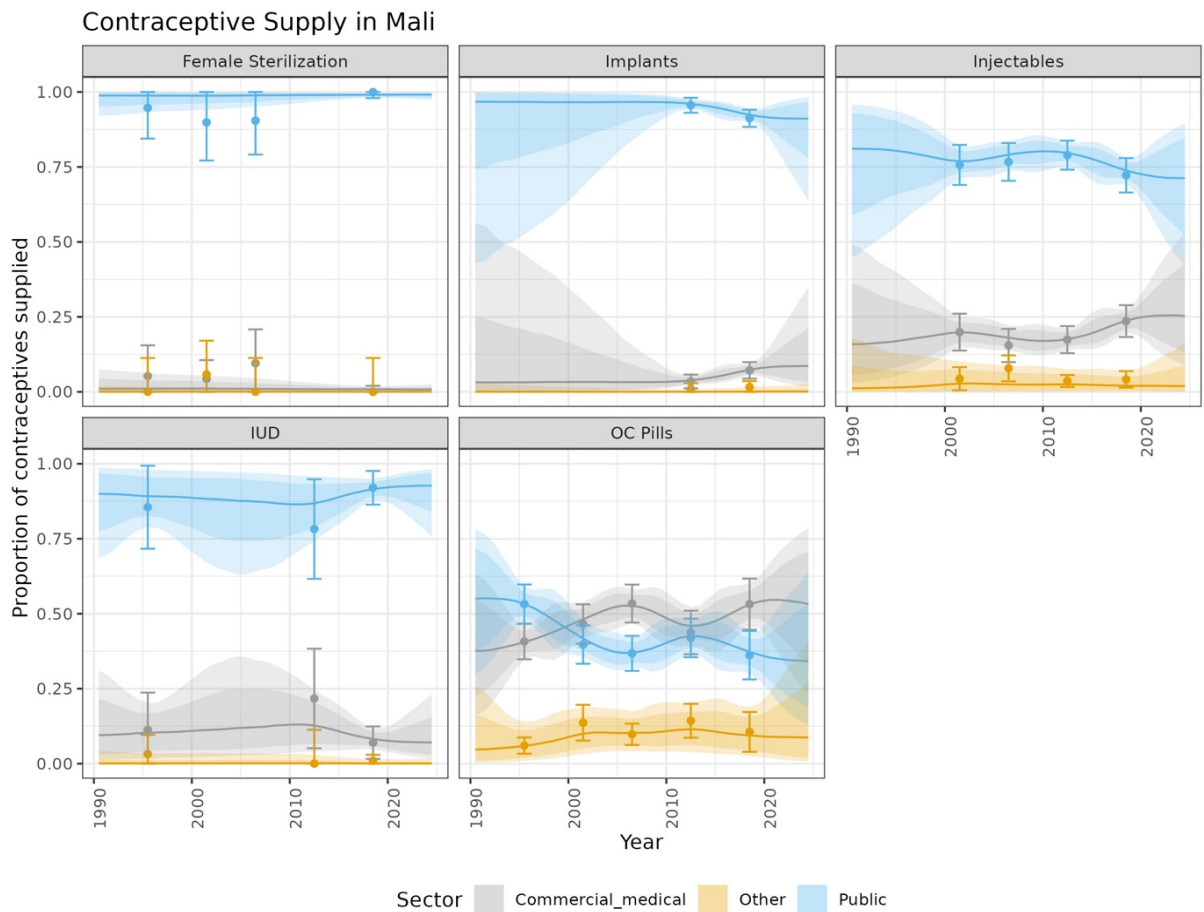

**Figure A.14.** The projections for the proportion of modern contraceptives supplied by each sector in Mali. The median estimates are shown by the continuous line while the 80% and 95% credible interval is marked by shaded coloured areas. The DHS data point is signified by a point on the graph with error bars displaying the standard error associated with each observation. The sectors are coloured blue for public, grey for commercial medical and gold for other private.

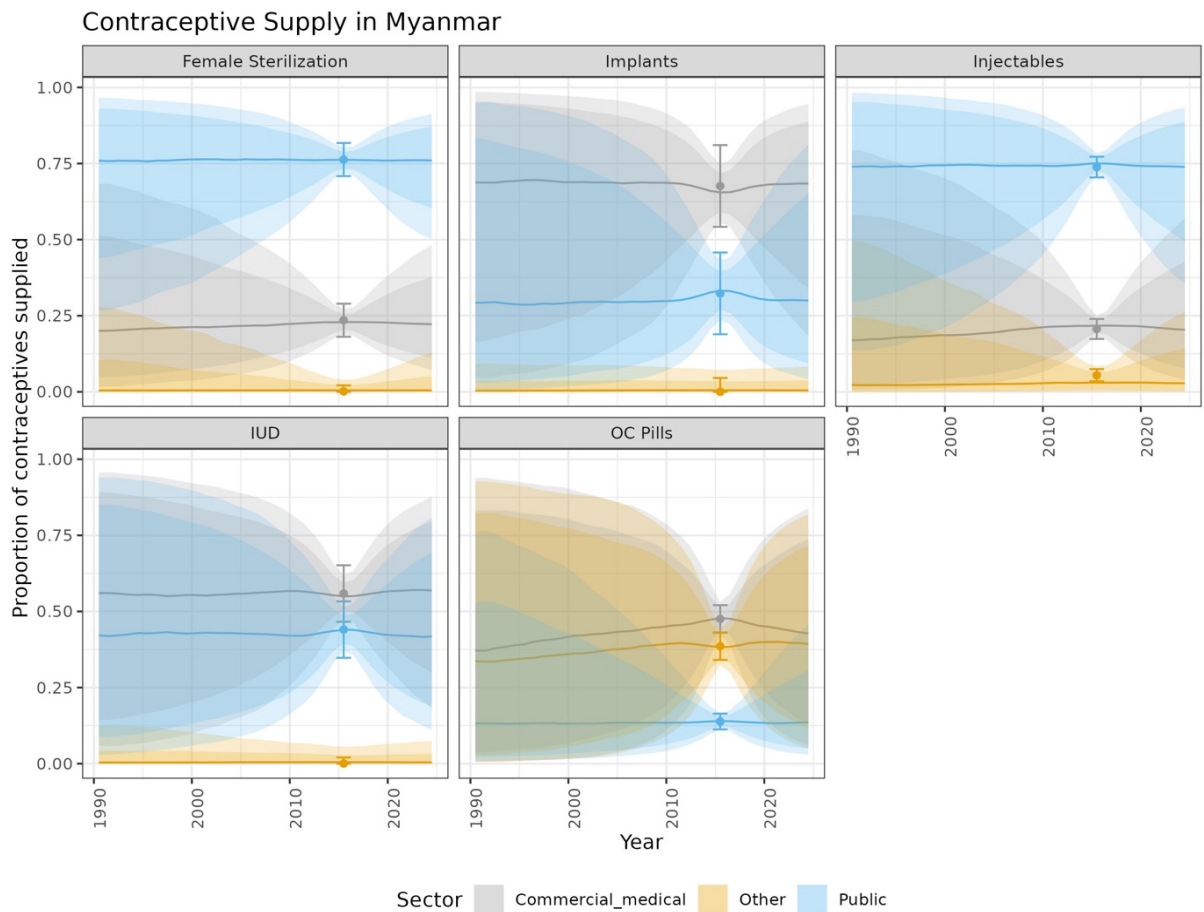

**Figure A.15.** The projections for the proportion of modern contraceptives supplied by each sector in Myanmar. The median estimates are shown by the continuous line while the 80% and 95% credible interval is marked by shaded coloured areas. The DHS data point is signified by a point on the graph with error bars displaying the standard error associated with each observation. The sectors are coloured blue for public, grey for commercial medical and gold for other private.

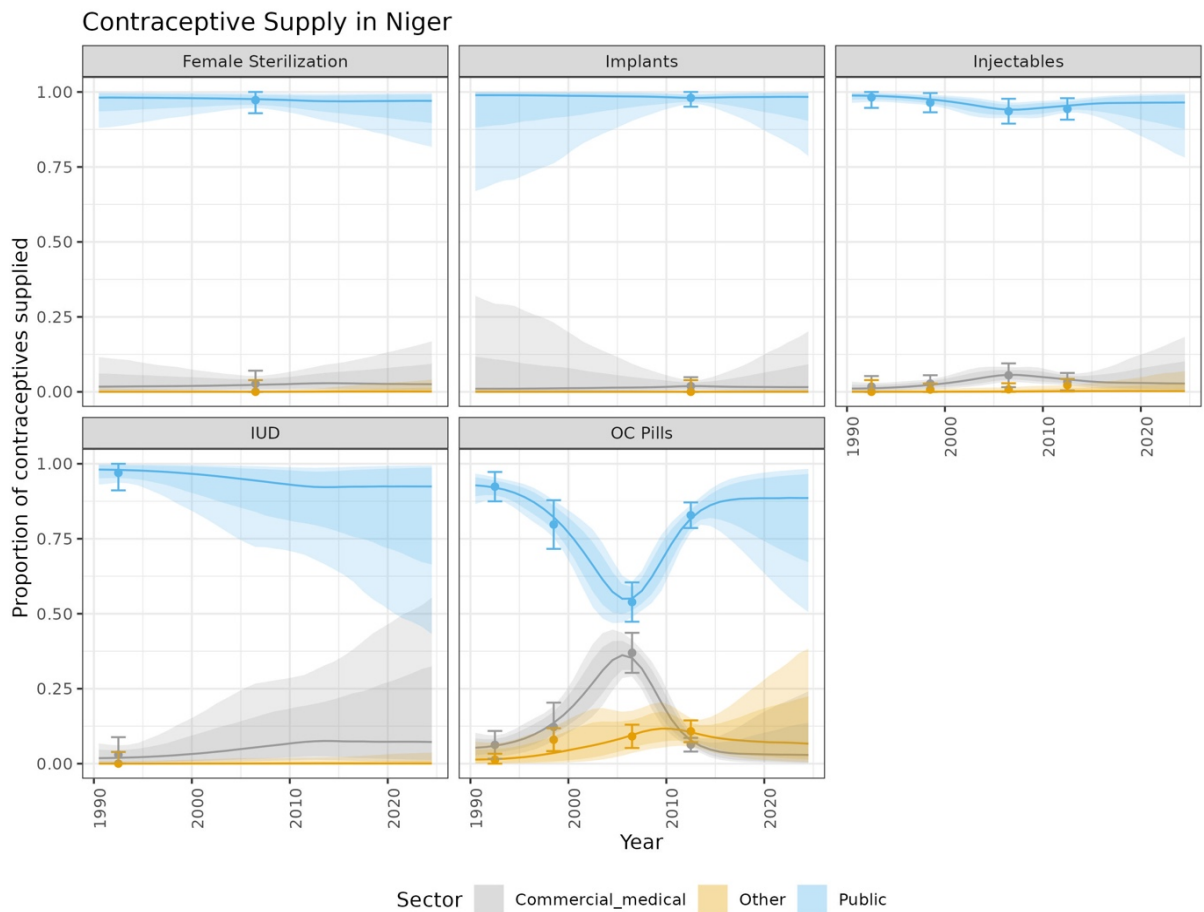

**Figure A.16.** The projections for the proportion of modern contraceptives supplied by each sector in Niger. The median estimates are shown by the continuous line while the 80% and 95% credible interval is marked by shaded coloured areas. The DHS data point is signified by a point on the graph with error bars displaying the standard error associated with each observation. The sectors are coloured blue for public, grey for commercial medical and gold for other private.

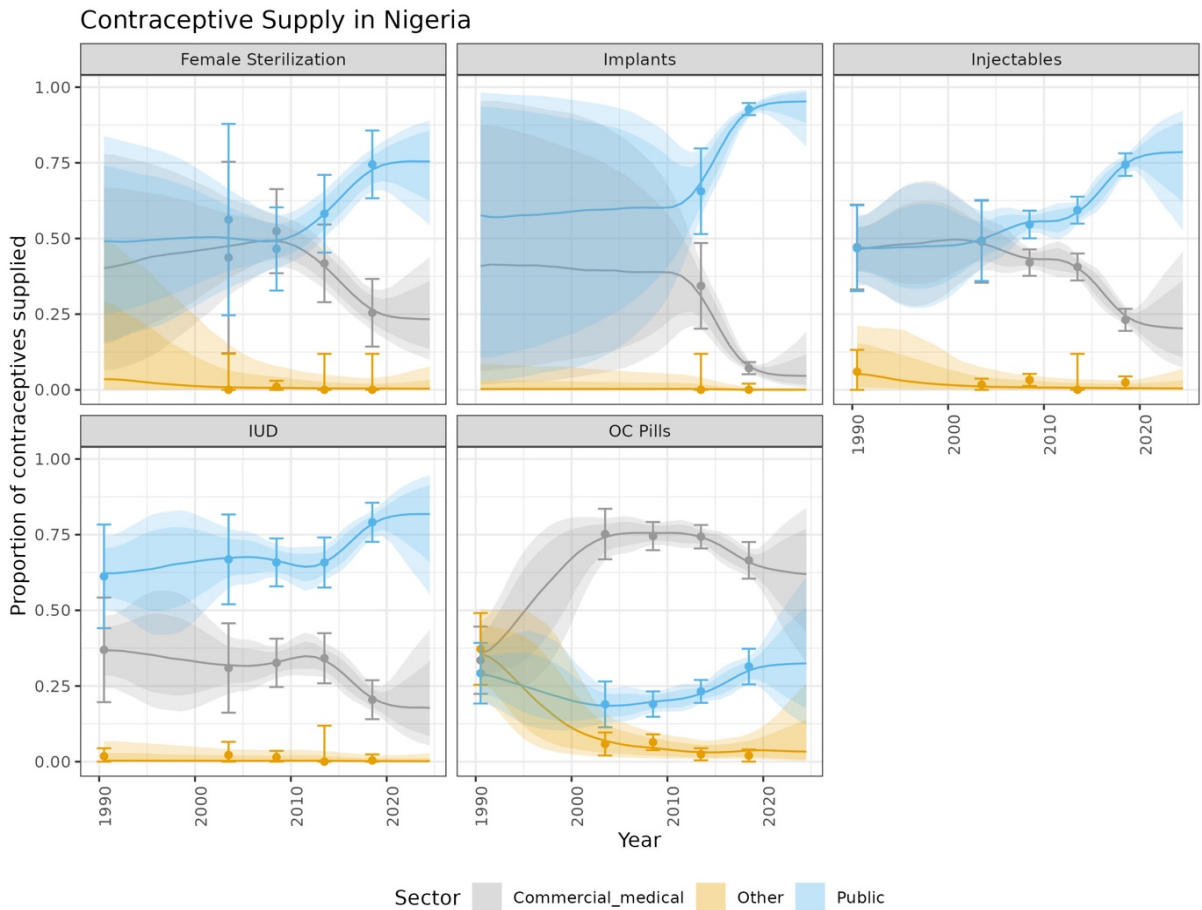

**Figure A.17.** The projections for the proportion of modern contraceptives supplied by each sector in Nigeria. The median estimates are shown by the continuous line while the 80% and 95% credible interval is marked by shaded coloured areas. The DHS data point is signified by a point on the graph with error bars displaying the standard error associated with each observation. The sectors are coloured blue for public, grey for commercial medical and gold for other private.

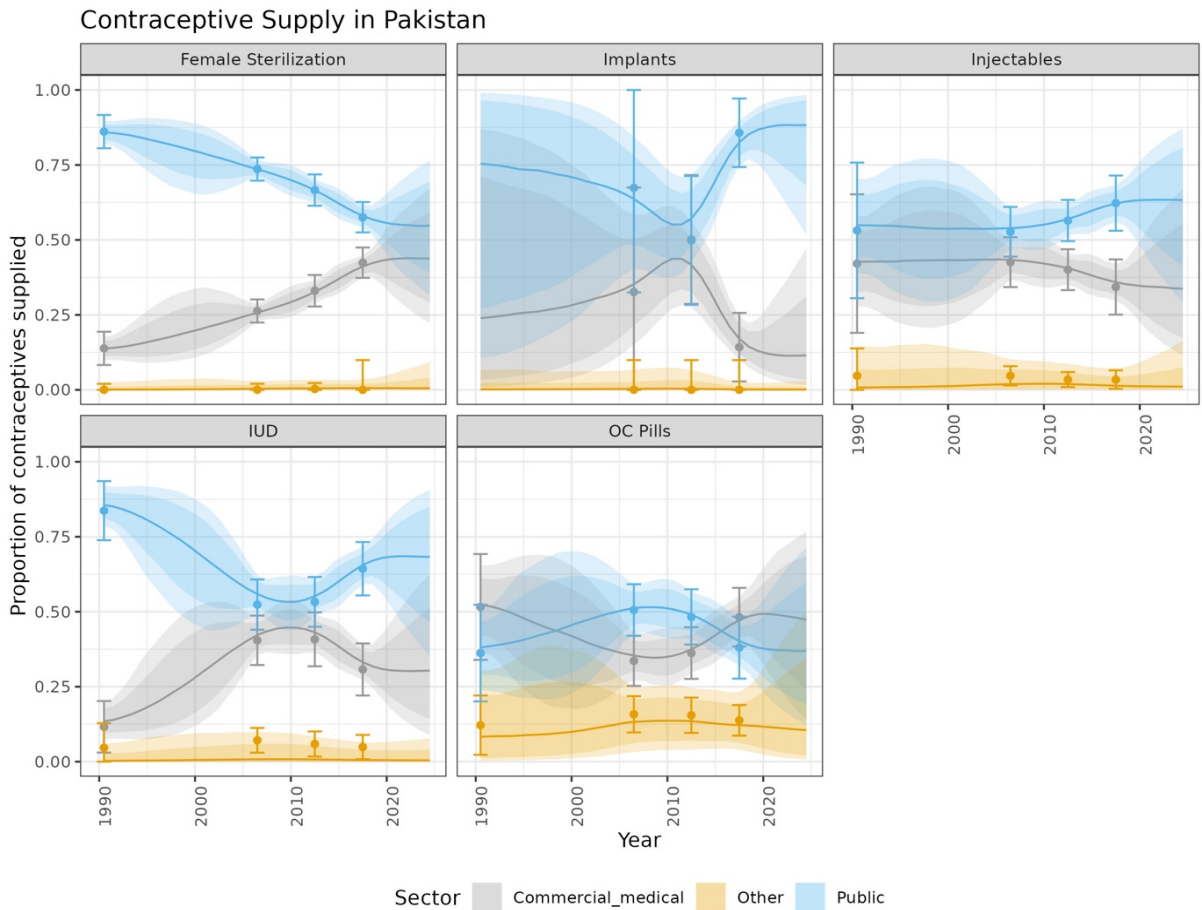

**Figure A.18.** The projections for the proportion of modern contraceptives supplied by each sector in Pakistan. The median estimates are shown by the continuous line while the 80% and 95% credible interval is marked by shaded coloured areas. The DHS data point is signified by a point on the graph with error bars displaying the standard error associated with each observation. The sectors are coloured blue for public, grey for commercial medical and gold for other private.

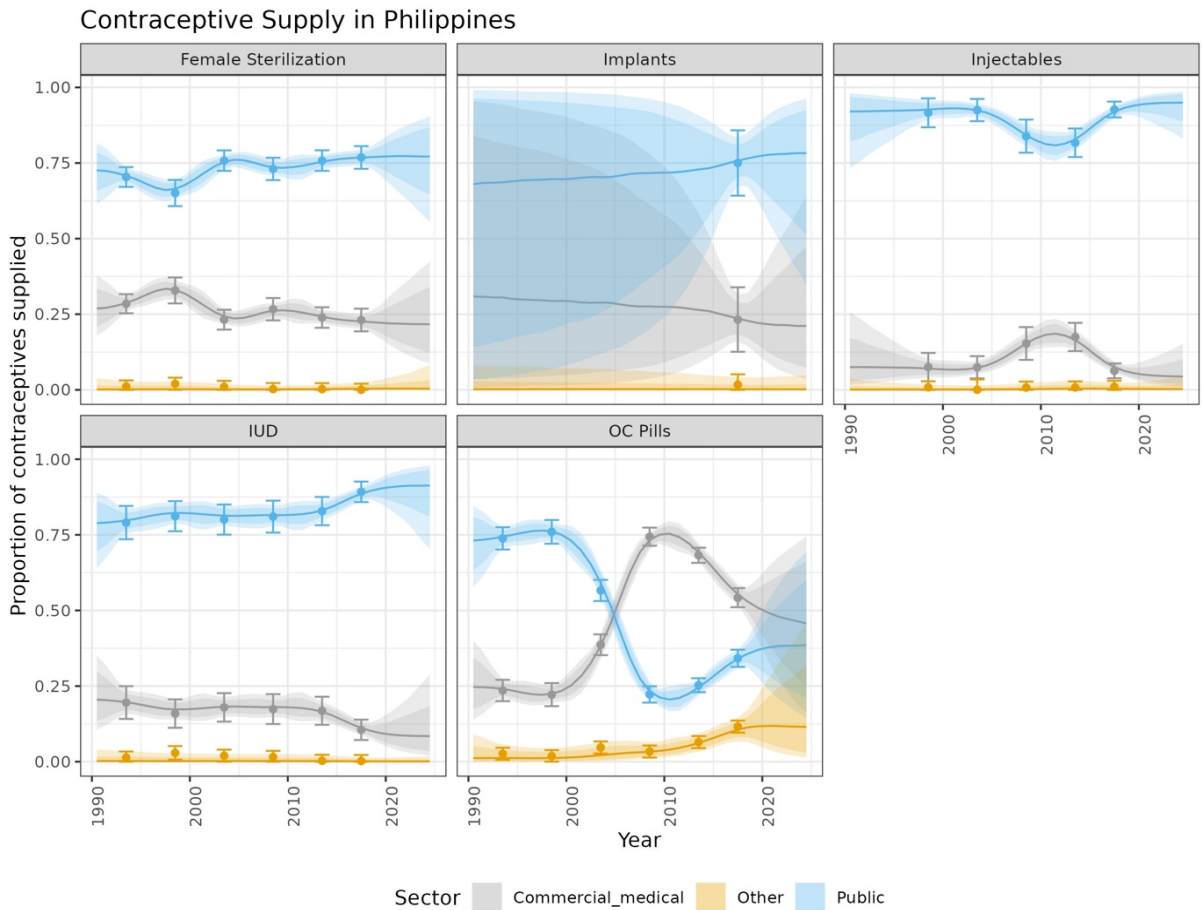

**Figure A.19.** The projections for the proportion of modern contraceptives supplied by each sector in the Philippines. The median estimates are shown by the continuous line while the 80% and 95% credible interval is marked by shaded coloured areas. The DHS data point is signified by a point on the graph with error bars displaying the standard error associated with each observation. The sectors are coloured blue for public, grey for commercial medical and gold for other private.

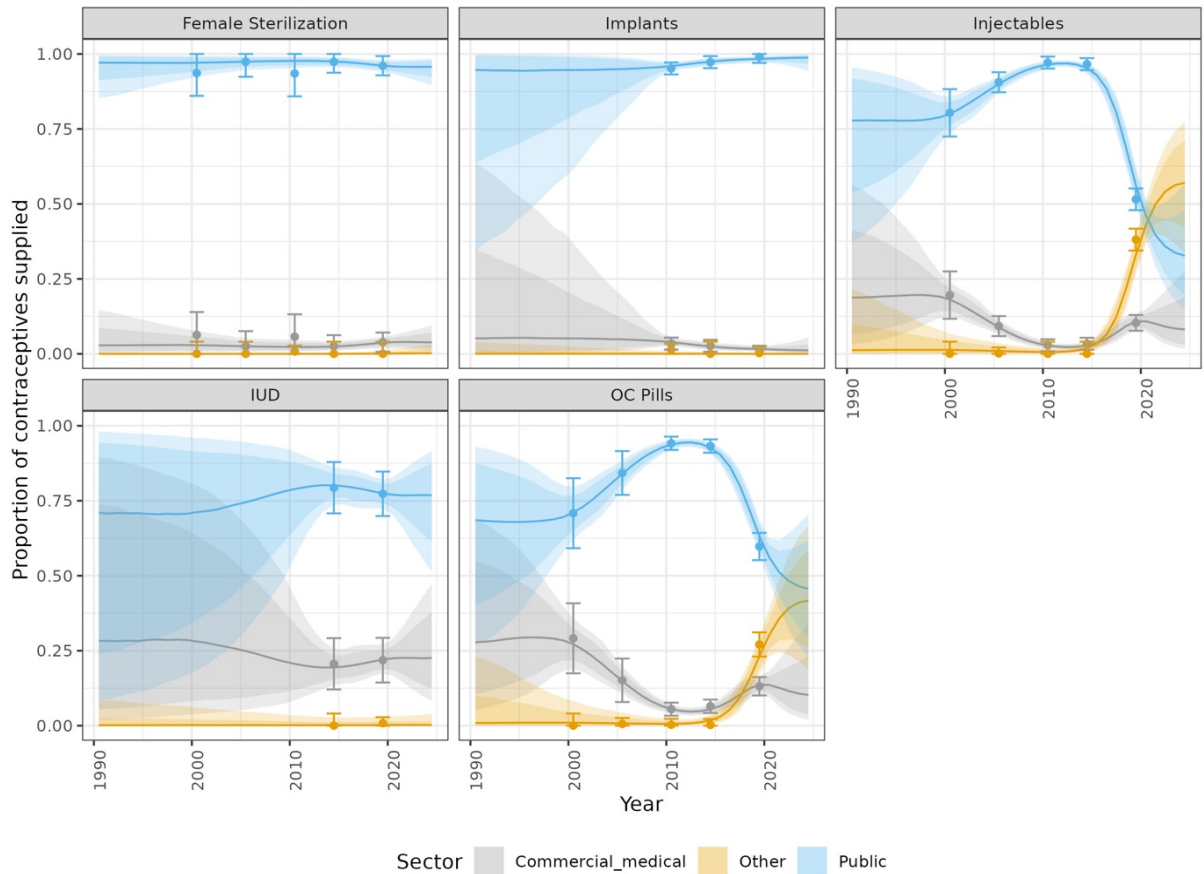

**Figure A.20.** The projections for the proportion of modern contraceptives supplied by each sector in Rwanda. The median estimates are shown by the continuous line while the 80% and 95% credible interval is marked by shaded coloured areas. The DHS data point is signified by a point on the graph with error bars displaying the standard error associated with each observation. The sectors are coloured blue for public, grey for commercial medical and gold for other private.

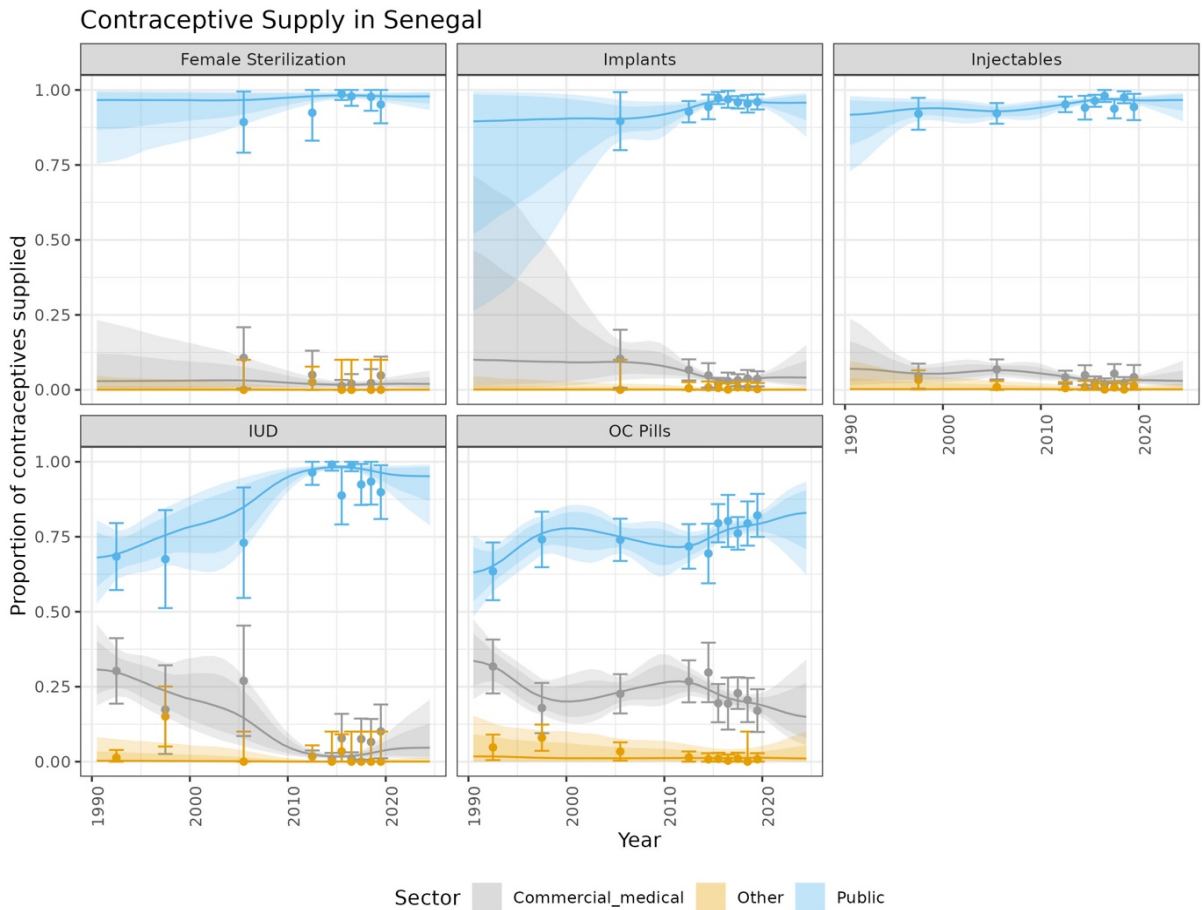

**Figure A.21.** The projections for the proportion of modern contraceptives supplied by each sector in Senegal. The median estimates are shown by the continuous line while the 80% and 95% credible interval is marked by shaded coloured areas. The DHS data point is signified by a point on the graph with error bars displaying the standard error associated with each observation. The sectors are coloured blue for public, grey for commercial medical and gold for other private.

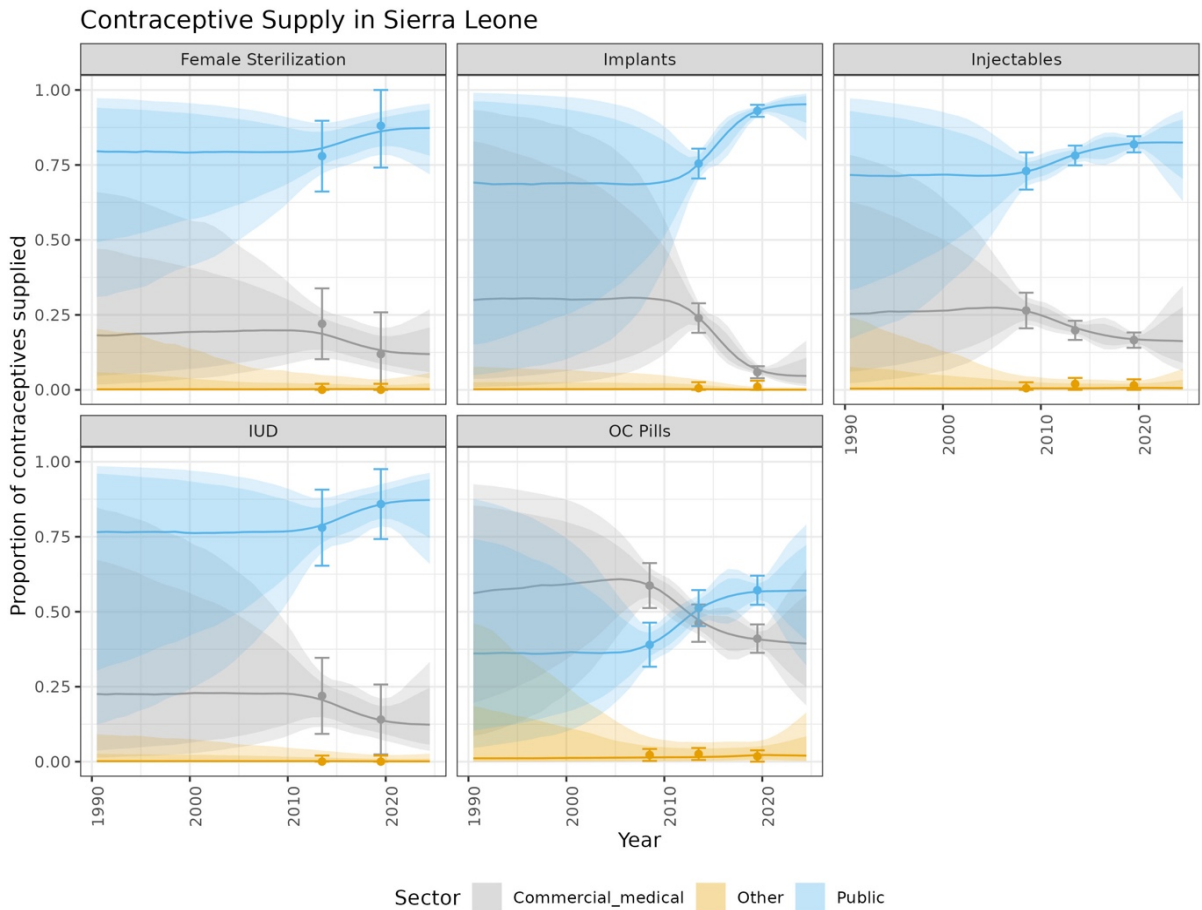

**Figure A.22.** The projections for the proportion of modern contraceptives supplied by each sector in Sierra Leone. The median estimates are shown by the continuous line while the 80% and 95% credible interval is marked by shaded coloured areas. The DHS data point is signified by a point on the graph with error bars displaying the standard error associated with each observation. The sectors are coloured blue for public, grey for commercial medical and gold for other private.

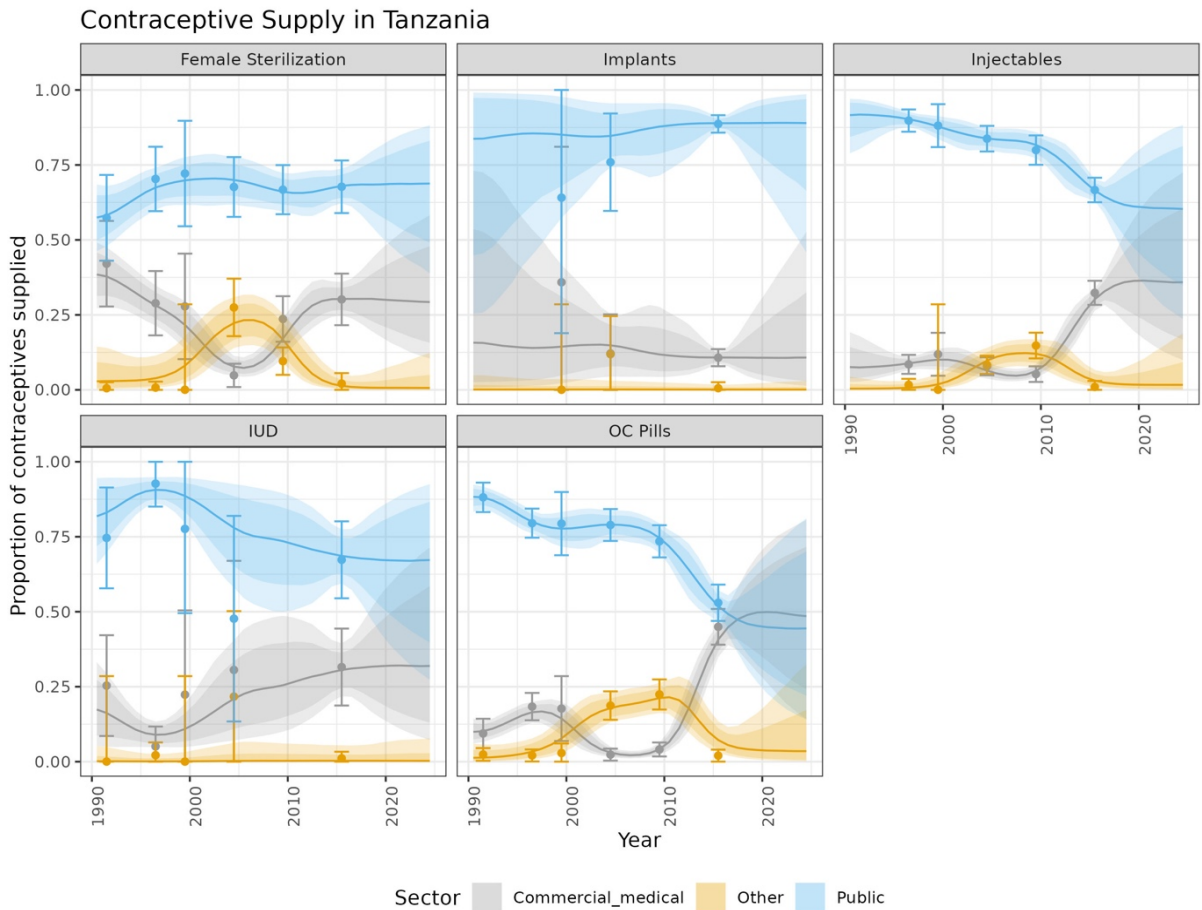

**Figure A.23.** The projections for the proportion of modern contraceptives supplied by each sector in Tanzania. The median estimates are shown by the continuous line while the 80% and 95% credible interval is marked by shaded coloured areas. The DHS data point is signified by a point on the graph with error bars displaying the standard error associated with each observation. The sectors are coloured blue for public, grey for commercial medical and gold for other private.

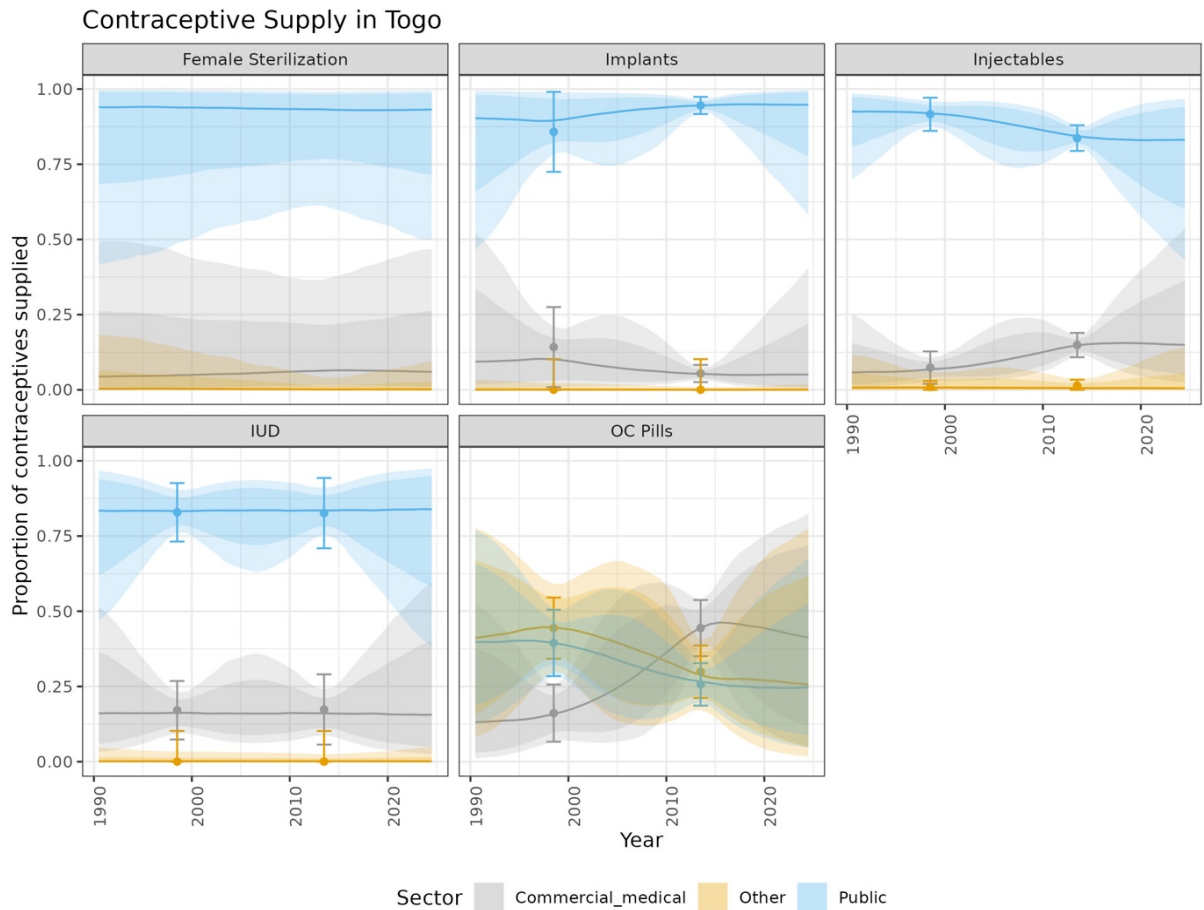

**Figure A.24.** The projections for the proportion of modern contraceptives supplied by each sector in Togo. The median estimates are shown by the continuous line while the 80% and 95% credible interval is marked by shaded coloured areas. The DHS data point is signified by a point on the graph with error bars displaying the standard error associated with each observation. The sectors are coloured blue for public, grey for commercial medical and gold for other private.

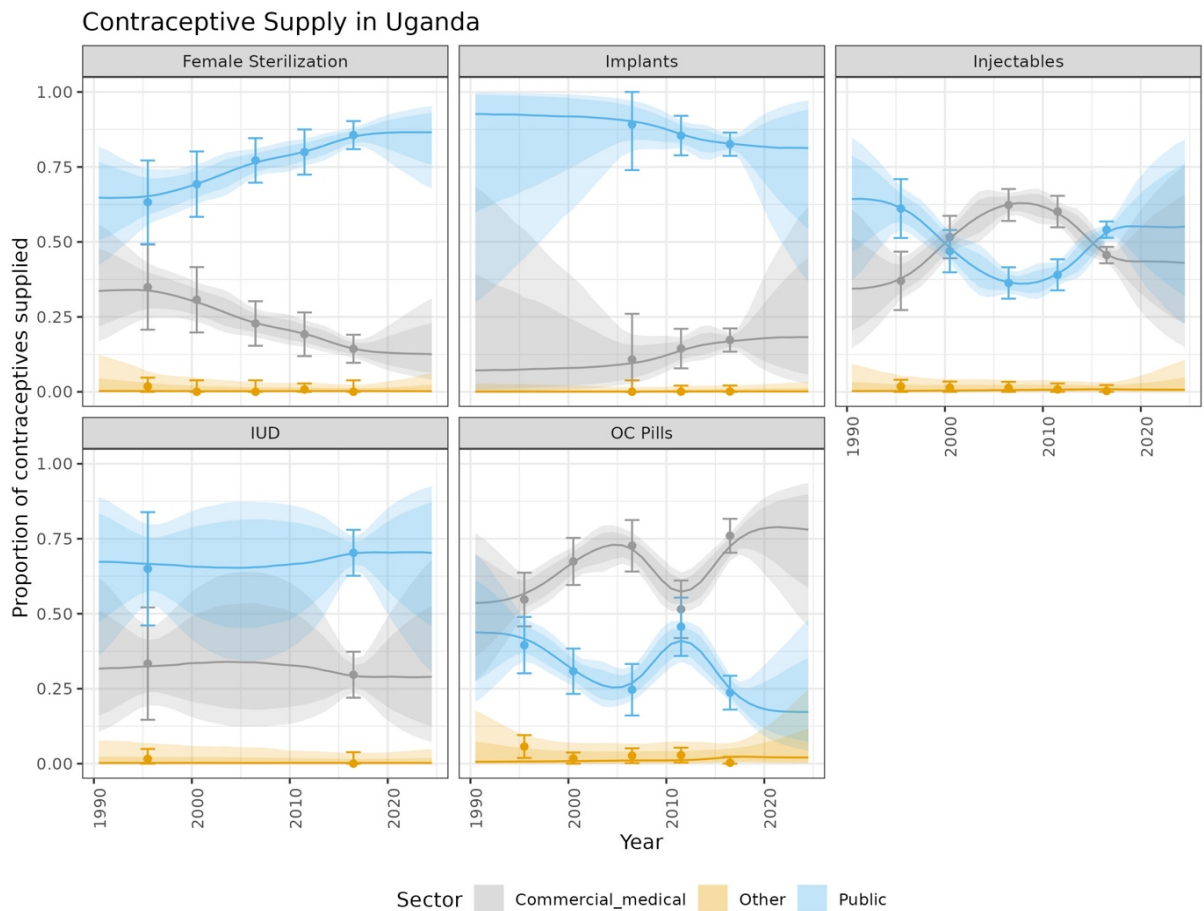

**Figure A.25.** The projections for the proportion of modern contraceptives supplied by each sector in Uganda. The median estimates are shown by the continuous line while the 80% and 95% credible interval is marked by shaded coloured areas. The DHS data point is signified by a point on the graph with error bars displaying the standard error associated with each observation. The sectors are coloured blue for public, grey for commercial medical and gold for other private.

**Calculating Estimated Modern Use**

At present, the method for calculating EMUs from service statistics is as follows[5]:

1. The raw data is reported. This data describes the numbers of family planning commodities distributed to clients, commodities distributed to health facilities, family planning facility visits, or family planning facility users.
2. Frequently, the raw data reported does not include the private sector. As such, the raw data may not be fully representative of the country's whole contraceptive market. To address this issue, the latest DHS survey of the country is used to provide a breakdown of the contraceptive market. These DHS quantities are used to scale up the data accordingly.

$$y_{adjusted} = \frac{y_{raw}}{p}$$

Where,

$y_{adjusted}$ is the adjusted number of family planning commodities /visits/users for a given contraceptive method, where the private sector is accounted for

$y_{raw}$ is the unadjusted number of family planning commodities /visits/users for a given contraceptive method, where the private sector data is unaccounted for

$p$ is the relevant latest DHS survey observation for proportion of a given contraceptive method supplied by the public/private commercial medical/private other sector.

3. Couple-year protection factors (CYPs) are applied to short-term contraceptive methods (STMs) to provide an estimate of the duration contraceptive protection supplied per unit of STM.
4. For LAPMs, we assume there exists a population of women who began using LAPMs before service statistic recording started. As such, a proportion of these women will continue to use their LAPMs into the study period of the service statistics. Historic users can be carried forward into recent years depending on estimated method continuation rates and CYP factor.
5. After these adjustments, the number of users of modern contraception is estimated.
6. Estimates of Modern use (EMU) is the proportion of women of reproductive age who are using modern contraception each year.

$$\text{EMU} = \frac{Estimated\ number\ of\ modern\ contraceptive\ users}{Total\ number\ of\ women\ of\ reproductive\ age} \quad (14)$$

An application of our model estimates is in the production of EMUs. We propose to use the model estimates in step 2 of the existing EMU methodology listed above. To do this, the full posterior sample for the contraceptive supply of sector s, method m, in country c, at time t is used to scale up the corresponding raw data. This provides a full posterior sample for the resulting steps and finally for the final EMUs.